\documentclass[%
 reprint,
 amsmath,amssymb,
 aps,
 prb,
 floatfix,
]{revtex4-2}
\usepackage{amsmath}
\usepackage{amssymb}
\usepackage[LGR,T1]{fontenc}
\usepackage[utf8]{inputenc}
\usepackage{graphicx}
\usepackage{subfig}
\usepackage{dcolumn}
\usepackage{bm}
\usepackage[unicode=true,pdfusetitle,
 bookmarks=true,bookmarksnumbered=false,bookmarksopen=false,
 breaklinks=true,pdfborder={0 0 0},pdfborderstyle={},backref=false,colorlinks=true]{hyperref}

\preprint{APS/123-QED}

\begin{document}

\title{Topological Winding Numbers from Wavefront Dislocations in Local Electronic Density}

\author{Yuval Abulafia}
\affiliation{Department of Physics, Technion -- Israel Institute of Technology, Haifa 3200003, Israel}

\author{Eric Akkermans}
\email{eric@physics.technion.ac.il}
\affiliation{Department of Physics, Technion -- Israel Institute of Technology, Haifa 3200003, Israel}

\date{\today}

\begin{abstract}
Topological materials are characterized by integer invariants that underpin robust quantized electronic properties, as exemplified by the Chern number in the integer quantum Hall effect. Yet, for most candidate systems, the observable linked to the topological invariant remains unknown, precluding direct verification of their topological nature.  We present a general method to identify topological materials by connecting the local electronic density~$\delta\rho(\bm{r})$ to Atiyah-Singer index theorems.  This method offers a concrete protocol for determining the winding number, the topological invariant associated with chiral-symmetric Hamiltonians. It also identifies a contour-independent wavefront dislocation pattern in $\delta\rho(\bm{r})$ arising from interference induced by topological defects and demonstrates its application to numerical simulations and to existing STM data.  The method clearly distinguishes topological states from non-topological ones through a unified, standardized filtering step, offering a definitive approach for identifying and characterizing quantum topological states and opening the door to their use as robust, entangleable building blocks in quantum technologies.
\end{abstract}

\maketitle

\section{Introduction}
Topological phases of matter~\cite{Hasan2010,Qi2011} have attracted broad attention owing to the existence of topological quantum states (TQS) that display robustness to disorder and unique many-body characteristics.  Certain TQS exhibit Coulomb-interaction responses akin to heavy-fermion $f$~states, as recently observed in magic-angle twisted bilayer graphene~\cite{Calderon2020,Song2022}.  These states can generate fractional charges~\cite{Jackiw2012,Sarma2015,Ovdat2020} or magnetic moments that may hybridize and support quantum entanglement.

A hallmark of nontrivial topology is the emergence of gapless edge states characterized by $\mathbb{Z}$ or $\mathbb{Z}_2$ invariants~\cite{Aidelsburger2015,Kane2005a,Kane2005b,Fu2006}.  However, the presence of edge states alone does not guaranty a topological origin.  Direct access to topological invariants is rare: the integer quantum Hall effect (IQHE) is a notable exception where the number of edge states equals the Chern number and is inferred from Hall conductivity~\cite{Klitzing1980,Thouless1982,Kohmoto1985}.  In most other platforms, whether quantum condensed matter, AMO, or classical physics setups, topological invariants are not measured directly~\cite{Aidelsburger2015,Bianco2011,Chalopin2020,Dutreix2017,Dutreix2021,Jotzu2014}, despite the proliferation of predicted and observed edge-like features.  The core difficulty is that topological signatures reside in wavefunctions, making their detection through transport or spectral methods challenging~\cite{Hashimoto2008}.  Indirect characterization via interfacing with selected materials~\cite{Bernevig2006,Koenig2007} can reveal edge states~\cite{Mancini2015,Hu2015,Mittal2016} but often falls short of establishing their topological nature.  
Closer in spirit to the present work, wavefront dislocations in Friedel oscillations have been used to determine the geometric Berry phase of the pristine graphene Dirac cone~\cite{Dutreix2019}. However, the Berry phase characterizes the band structure of the unperturbed crystal and equals ~$\pi$ in graphene, irrespective of any local perturbation; it carries no information about whether a given defect induces a topological zero mode or not.
The winding number measured here is instead a property of the full Hamiltonian, including the defect; it counts the number of topological zero modes via the index theorem and vanishes for symmetry-breaking perturbations, such as an adatom, making it a sharper and more informative topological diagnostic than the Berry phase.

We adopt the tenfold classification of insulators and superconductors~\cite{Kitaev2009,Schnyder2008,Altland1997}, relying on two anti-unitary symmetries (time-reversal and particle-hole), a unitary chiral symmetry, and spatial dimension~$d$.  Although initially confined to weakly interacting fermions in periodic potentials, this classification has been extended to include defects~\cite{Goft2023,Teo2010,Chiu2016}, enabling TQS to be induced by specifically designed defects~\cite{Goft2023,Goft2025,Abulafia2025}.

In this work, we propose a systematic identification of TQS through an interference scheme for scanning tunneling microscopy (STM), combined with index theorems~\cite{Atiyah1968,Nakahara1990,Niemi1984,Atiyah1963,Yankowsky2013}, which is a powerful mathematical framework.  The Atiyah-Singer index theorems (ASIT) identify topological edge states, also called zero modes, and determine the relevant integer invariants.  We connect these integers to variations in the local electronic density~$\delta\rho(\bm{r})$ imaged by STM, providing a clear and decisive identification of TQS.  
A companion work~\cite{Abulafia2025} builds on the present results to derive systematic methods for constructing defect potentials that generate specific topological invariants and to establish their continuum-limit description. The present work focuses on the identification and measurement protocol: how topological invariants are extracted from density images through a concrete filtering and winding-number evaluation procedure, and how this protocol discriminates topological from non-topological perturbations.
The new results are: (i)~the zero-mode dominance equation connecting $\delta\rho(\bm{r})$ mechanistically to the topological wavefunctions via chiral symmetry; (ii)~an explicit, step-by-step Fourier-filtering protocol for extracting the topological invariant from STM or numerical data; (iii)~a quantitative, contour-independent verification of the winding number $\nu = \pm 1$ on the graphene vacancy; and (iv)~a deliberate non-topological counterexample (adatom) demonstrating that the method discriminates rather than merely detects topology.
\section{Formalism}
\subsection{Index theorems and chiral Hamiltonians} ASIT connects spectral properties of specific elliptic operators to topological numbers computed from a Weyl transform of their symbol~\cite{Goft2023}.  For chiral Hamiltonians,
\begin{equation}
H = \begin{pmatrix} 0 & \mathcal{D}^\dagger \\ \mathcal{D} & 0 \end{pmatrix},
\label{eq:chiral_H}
\end{equation}
the analytical index, $\mathrm{Index}\,\mathcal{D} \equiv \dim\ker\mathcal{D} - \dim\ker\mathcal{D}^\dagger$, is a finite integer related by the index theorem to a bulk topological number~$\nu$: $\mathrm{Index}\,\mathcal{D} = \nu$.  Having $\mathcal{D} = \mathcal{D}^\dagger$ implies a vanishing index and hence the absence of TQS.  A nonzero index signals an imbalance in zero modes between $\mathcal{D}$ and $\mathcal{D}^\dagger$.  These zero modes need not correspond to zero-energy states of $H$; crucially, however, both $\mathrm{Index}\,\mathcal{D}$ and $\nu$ enumerate the number~$N_{\mathrm{zm}}$ of TQS.  The topological integer~$\nu$ either a Chern or a winding number, is connected to nontrivial $\mathbb{Z}$ or $\mathbb{Z}_2$ classes.  We thus obtain the chain of equalities,
\begin{equation}
N_{\mathrm{zm}} = |\mathrm{Index}\,\mathcal{D}| = |\nu|.
\label{eq:chain}
\end{equation}
This equation provides the framework to construct, recognize, and characterize TQS.  The operator~$\mathcal{D}$ is the first building block for establishing topology; designing topological materials requires creating specific potentials~\cite{Abulafia2025} that shape~$\mathcal{D}$.  The equality $N_{\mathrm{zm}} = |\nu|$ expresses bulk-edge correspondence between the number of topological edge states and the bulk invariant.

\subsection{Winding number from local density} For chiral Hamiltonians~\eqref{eq:chiral_H} with a localized potential, $\nu$ can be read off STM variations in local electronic density~\cite{Abulafia2025},
\begin{equation}
\delta\rho(\bm{r}) = -\mathrm{Re}\!\left(iF(r)\sin\theta\, e^{i\chi(\bm{r})}\right),
\label{eq:drho}
\end{equation}
via the winding number of the phase $\chi(\bm{r})$,
\begin{equation}
\frac{1}{2\pi}\oint_{\mathcal{C}} \nabla\chi(\bm{r})\cdot d\bm{r} = \nu \in \mathbb{Z},
\label{eq:winding}
\end{equation}
for \emph{any} contour~$\mathcal{C}$ encircling the perturbation. Thus, measurement of~$\nu$ reduces to identifying an interference pattern in $\delta\rho(\bm{r})$ and evaluating its winding.  Although topological numbers are tied to the band structure, they cannot be directly inferred from the energy spectrum; instead, they emerge from the local wavefunction amplitudes,
\begin{equation}
\rho(\bm{r}) = -\frac{1}{\pi}\int dE\,\mathrm{Im}\sum_n \frac{|\varphi_n(\bm{r})|^2}{E - E_n + i0^+},
\label{eq:rho}
\end{equation}
accessible through STM imaging.  The key mechanism connecting $\delta\rho(\bm{r})$ to topology is a consequence of chiral symmetry: all nonzero-energy eigenstates $\varphi_{n}\left(\boldsymbol{r}\right)$ of~\eqref{eq:chiral_H} appear in pairs $\pm E_n$, so their contributions to $\delta\rho$ cancel pairwise.  The density variation is therefore entirely determined by the zero modes~\cite{Faktor2026, Ovdat2020}, as derived in Appendix.\ref{appendix:zero_mode_eq}
\begin{equation}
\delta\rho(\bm{r}) = \frac{e}{2}\sum_n \mathrm{sign}(E_n)\,|\varphi_n(\bm{r})|^2,
\label{eq:drho_zeromodes}
\end{equation}
and the dislocation in $\delta\rho(\bm{r})$ is a direct imprint of the phase winding of the topological zero-mode wavefunction.  STM thus provides access to the zero modes without resolving individual wavefunctions.

\subsection{Chiral phases and fermion doubling} In condensed matter, chiral phases arise on non-Bravais bipartite lattices that support extra pseudospin degrees of freedom, appearing as pairs of degenerate points $(K,K')$ in the Brillouin zone where the gap vanishes. The low-energy behavior is governed by a Dirac-like equation; hence, $(K,K')$ are known as Dirac points or valleys.  The existence of valley pairs (fermion doubling) is a universal consequence of the Nielsen-Ninomiya no-go theorem~\cite{Nielsen1983}, which requires extra fermion species for massless chiral fermions on a lattice.  For lattice Hamiltonians, the topological requirement of two valleys is independent of perturbations, so long as they preserve chirality and massless fermions~\cite{Katsnelson1979}. Graphene realizes precisely this setting.

\section{Application to graphene} We consider a honeycomb lattice, representative of graphene in the nearest-neighbor tight-binding approximation. The two Dirac points $(K,K')$ carry opposite winding numbers, yielding a non-topological phase.  We then insert a spatially localized potential coupling the two valleys, tuned to either preserve or break chiral symmetry.

\begin{figure}[t]
  \centering
  \subfloat[]{\includegraphics[width=0.51\columnwidth]{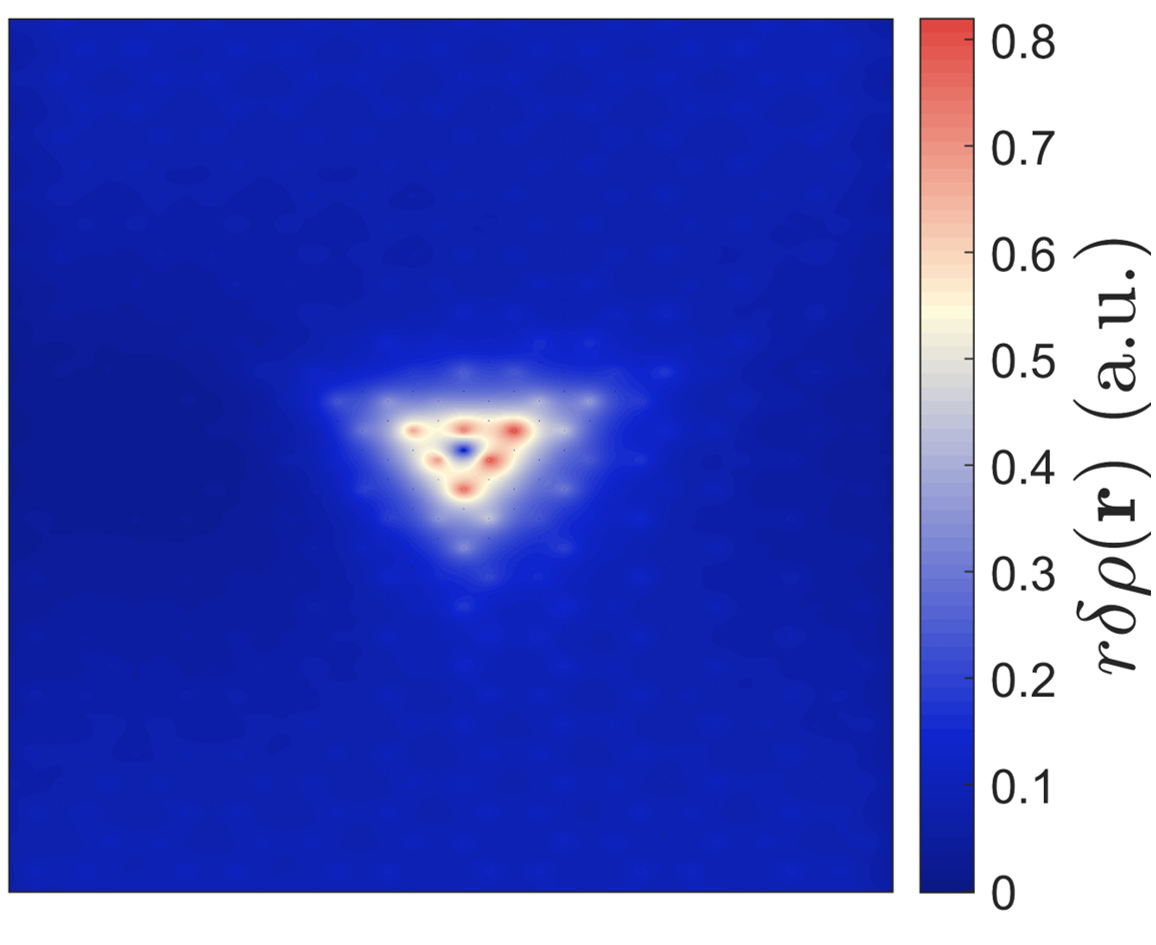}}
  \hfill
  \subfloat[]{\includegraphics[width=0.465\columnwidth]{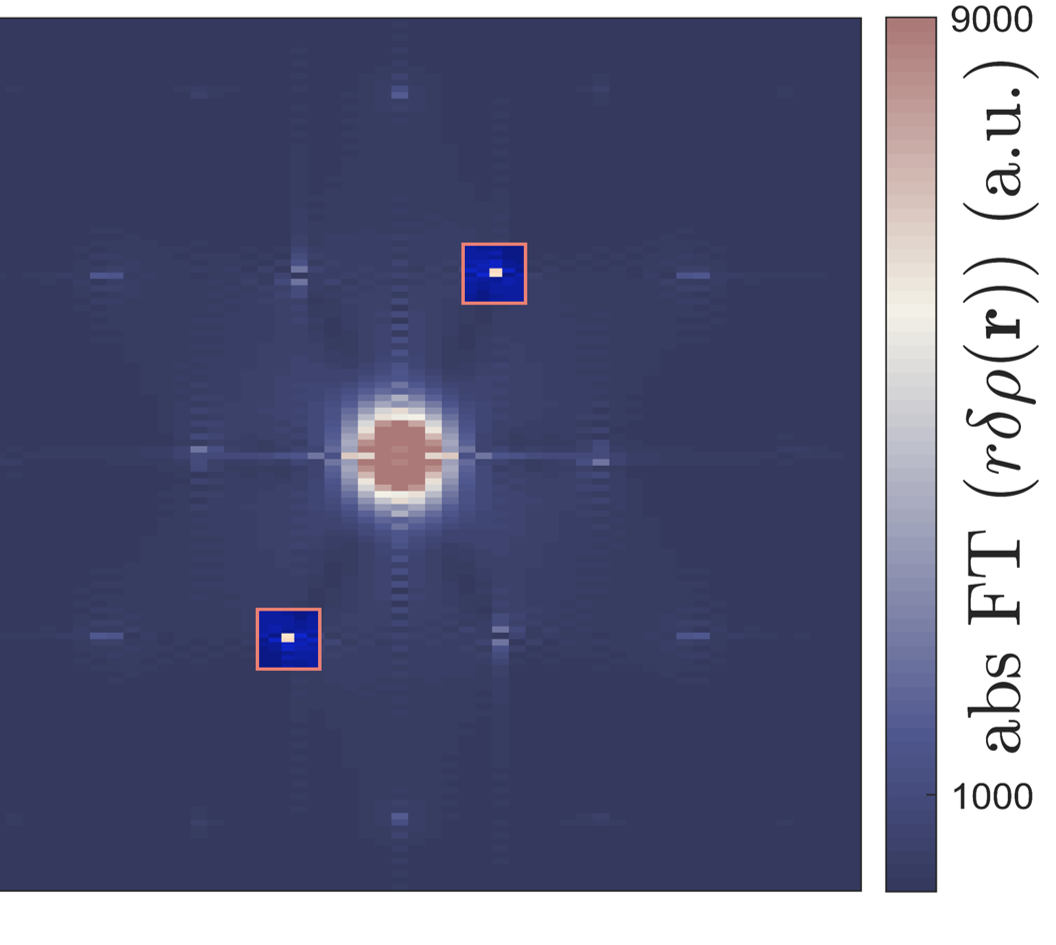}}\\
  \subfloat[]{\includegraphics[width=0.51\columnwidth]{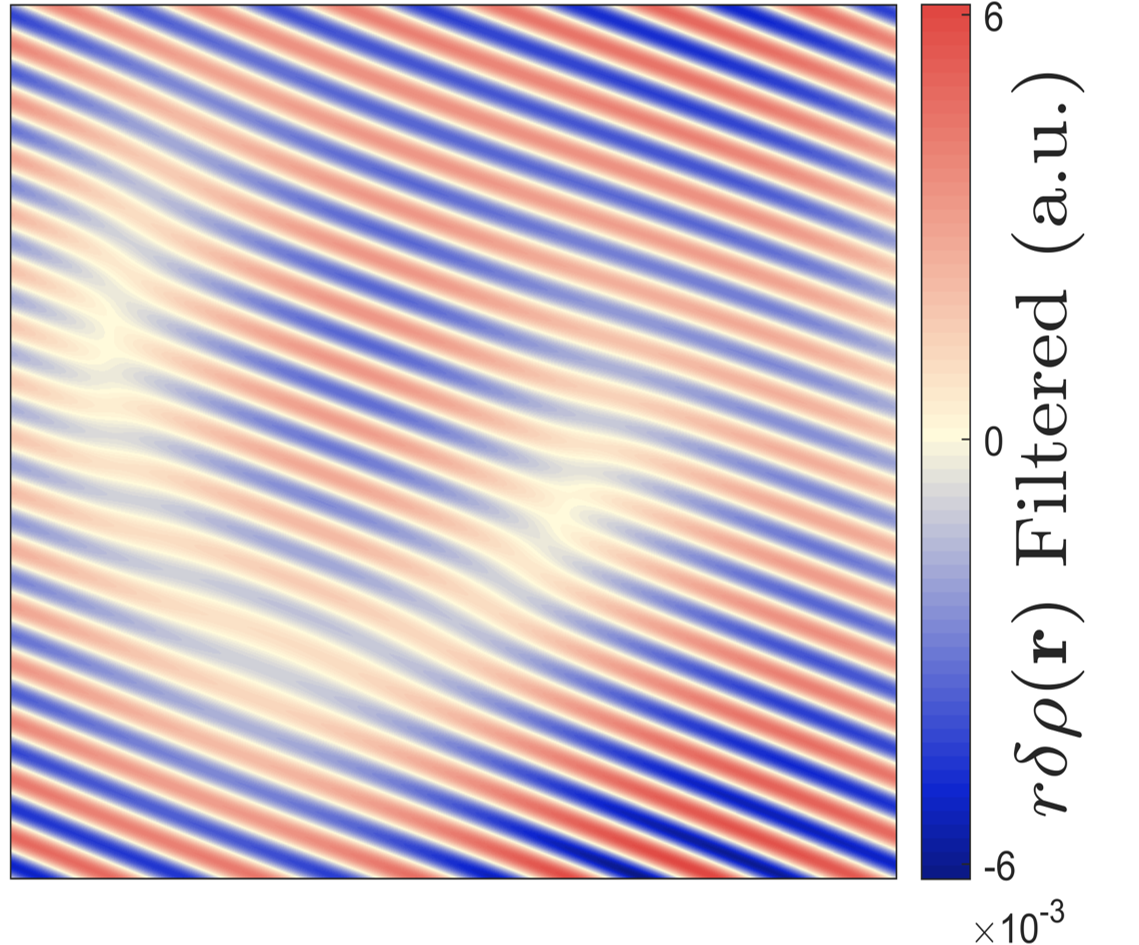}}
  \caption{Honeycomb lattice with a vacancy at the origin.
  (a)~Numerical $r\delta\rho(\bm{r})$ for a $36\times57$ tight-binding model with periodic boundary conditions; a $36\times30$ region surrounding the vacancy is shown (see Fig.~\ref{fig:sparial_behavior_supp}d).
  (b)~$|\mathrm{FT}(r\delta\rho)|$; the inset shows six satellite points from three intervalley pairs (one pair marked by red squares). Two diametrically opposite points are filtered.
  (c)~Inverse FT of the filtered data with restored phase, revealing a clear wavefront dislocation.}
  \label{fig:vacancy}
\end{figure}

\begin{figure}[t]
  \centering
 \subfloat[ ]
 {\includegraphics[width =0.45\columnwidth,height=3cm]{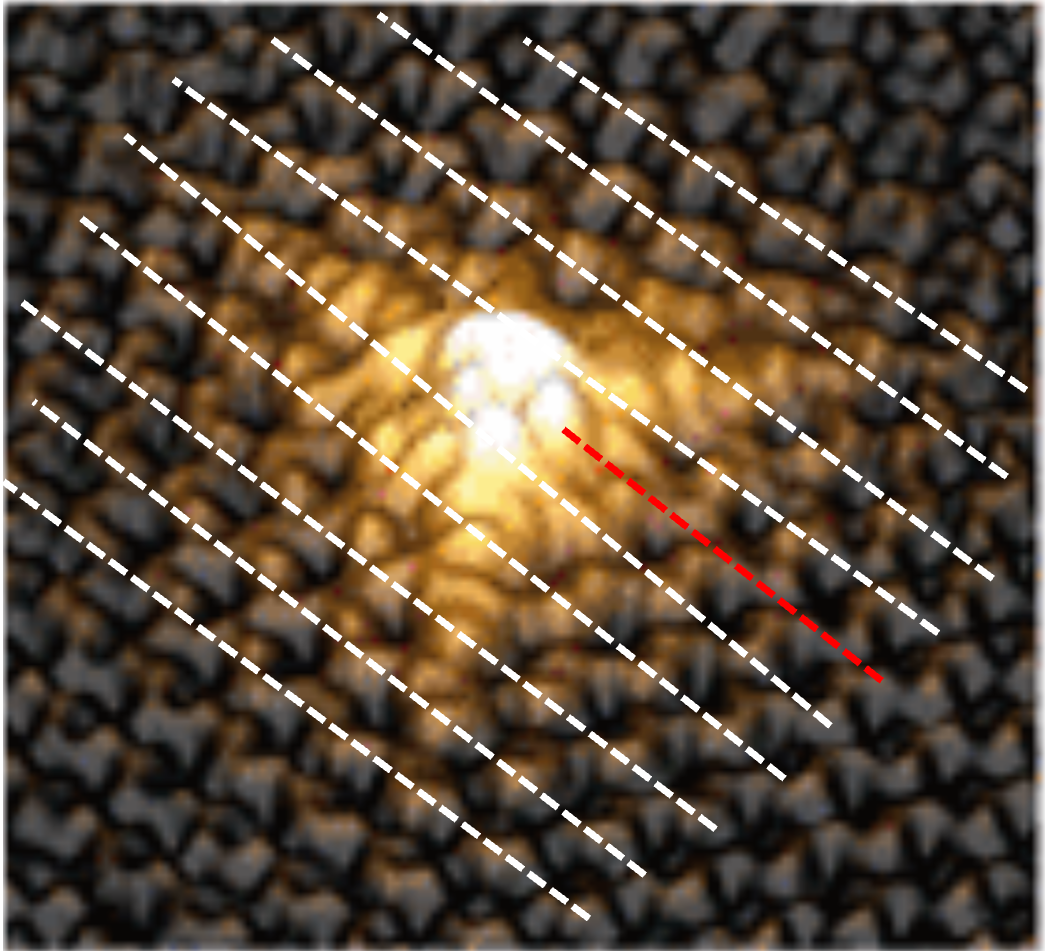}}
 \subfloat[ ]
 {\includegraphics[width =0.45\columnwidth,bb=40bp 40bp 350bp 275bp,clip]{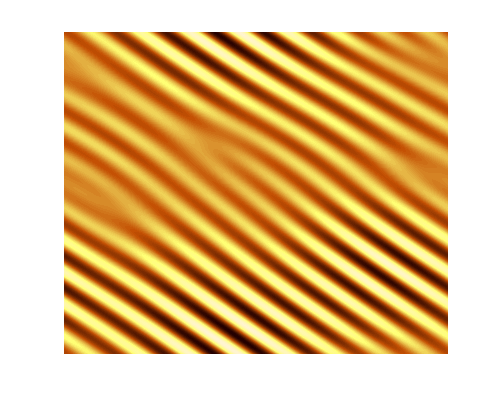}}
  \caption{STM measurements of graphene with a vacancy~\cite{Ugeda2010}.
  (a)~Raw data with added wavefront guides (dashed lines); one wavefront terminates at the vacancy.
  (b)~After FT filtering to two intervalley contributions, clearly showing one wavefront dislocation.}
  \label{fig:STM}
\end{figure}

For the chirality-preserving \textit{vacancy} potential, the intervalley coupling lifts the zero total winding number, rendering the perturbed lattice topological \cite{Goft2023, Abulafia2025}.  This is revealed by a distinct interference pattern in $\delta\rho(\bm{r})$, computed numerically (Fig.~\ref{fig:vacancy}a) and observed experimentally~\cite{Ugeda2010} (Fig.~\ref{fig:STM}a).  We extract the topological signal through the following protocol.  First, we form $r\delta\rho(\bm{r})$: since the relevant oscillations decay as $1/r$, this product tends to a constant amplitude at large $r$, concentrating the spectral weight in Fourier space into sharp, well-localized peaks rather than diffuse rings.  We then compute $\mathrm{FT}(r\delta\rho)$; its magnitude (Fig.~\ref{fig:vacancy}b) reveals three pairs of diametrically opposed satellite points reflecting the honeycomb symmetry—a direct manifestation of fermion doubling.  We select one such pair of peaks and retain the \emph{full complex} Fourier values, amplitude and phase together, within a window centered on those two points, setting all other Fourier components, including the dc term, to zero.  The inverse FT of this complex filtered spectrum directly yields the dislocation image (Fig.~\ref{fig:vacancy}c), in which a clear \emph{wavefront dislocation} is visible.  The same protocol is applied identically to the adatom case; the presence or absence of a dislocation in the result is therefore a direct, protocol-independent diagnostic of the topological character of the perturbation.

\begin{figure}[t]
  \centering
  \includegraphics[width=\columnwidth]{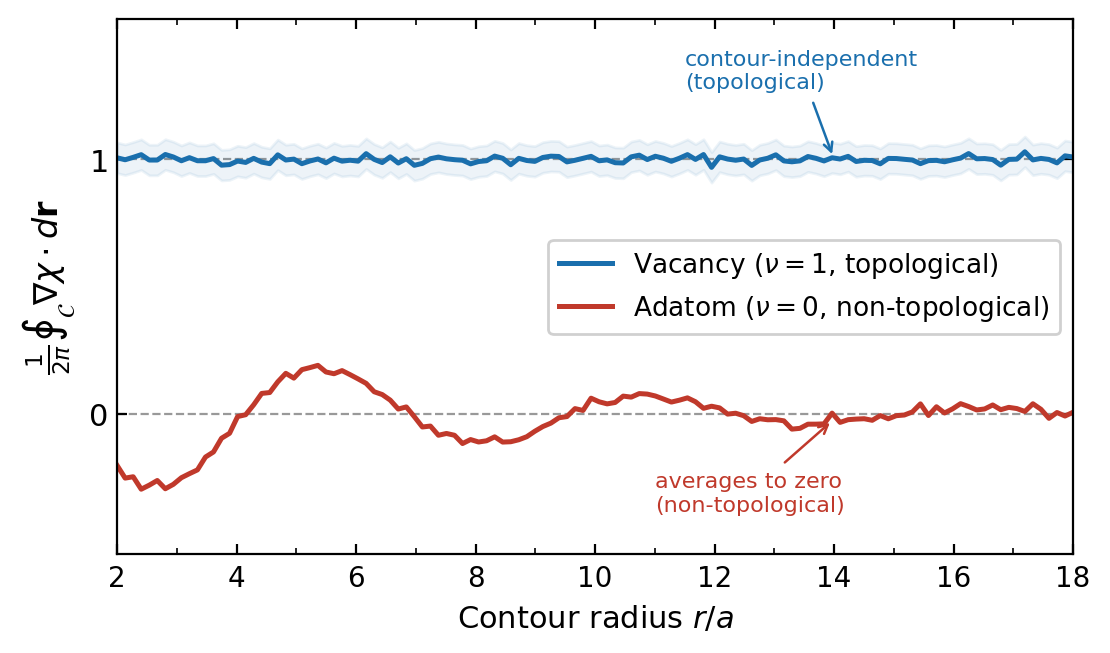}
  \caption{Winding-number contour integral $(2\pi)^{-1}\oint_{\mathcal{C}}\nabla\chi\cdot d\bm{r}$ as a function of contour radius~$r/a$, evaluated numerically on the filtered tight-binding densities for both defect types.  The vacancy (blue) yields $\nu = 1$ independent of contour size, the hallmark of a genuine topological invariant.  The adatom (red) yields a winding that decays to zero at large~$r$, confirming the absence of a global topological invariant and demonstrating that the same protocol unambiguously discriminates the two cases.}
  \label{fig:winding}
\end{figure}

\begin{figure}[t]
  \centering
  \subfloat[]{\includegraphics[width=0.48\columnwidth]{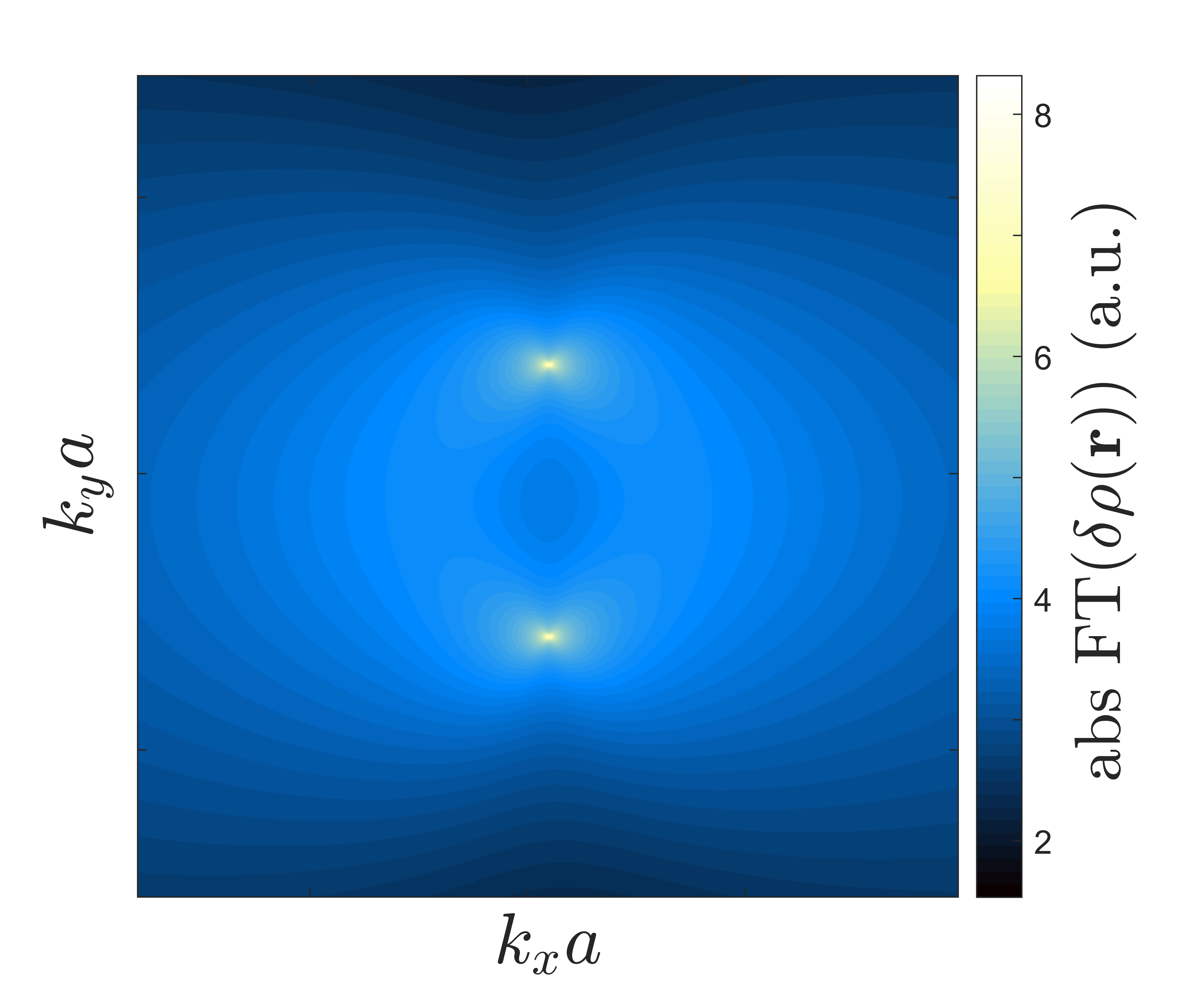}}
  \hfill
  \subfloat[]{\includegraphics[width=0.48\columnwidth]{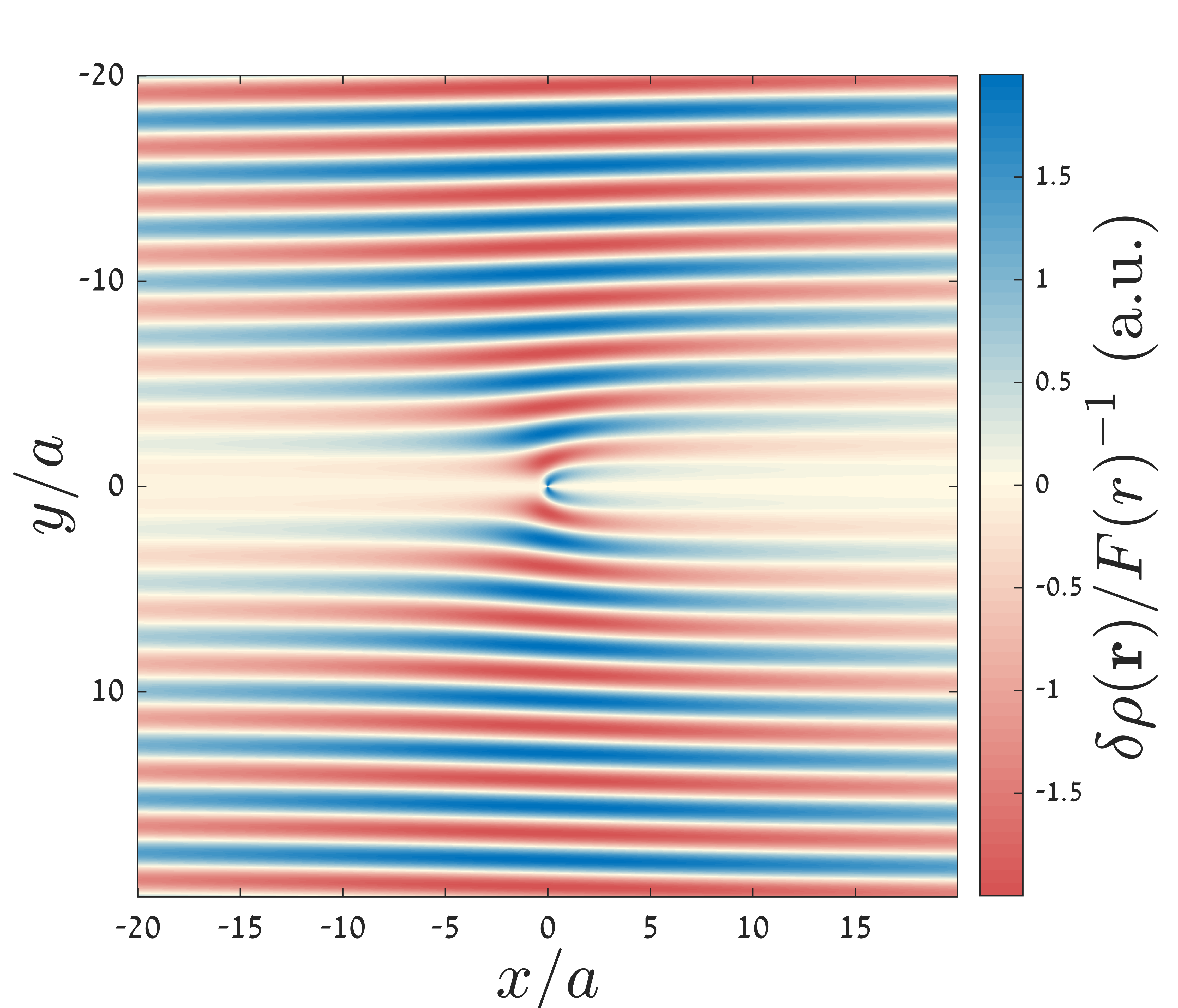}}
  \caption{Intervalley density variation from Eq.~\eqref{eq:drho_vacancy} for a vacancy.
  (a)~$|\mathrm{FT}(\delta\rho)|$ with $F(r) = r^{-2}$, showing two fermion-doubling peaks.
  (b)~Real-space plot exhibiting a single wavefront dislocation.}
  \label{fig:analytical}
\end{figure}

The intervalley contribution to $\delta\rho(\bm{r})$ for a vacancy reads~\cite{Abulafia2025}
\begin{equation}
\delta\rho_{\Delta K}(\bm{r}) = F(r)\left[\cos\chi(\bm{r},0) - \cos\chi(\bm{r},2)\right],
\label{eq:drho_vacancy}
\end{equation}
where $\bm{r} = (r,\theta)$ is measured from the vacancy, $F(r)$ decays at infinity, $\Delta\bm{K} \equiv \bm{K} - \bm{K}'$ connects the two valleys, and
\begin{equation}
\chi(\bm{r},n) \equiv \Delta\bm{K}\cdot\bm{r} + n\theta.
\label{eq:chi}
\end{equation}
Expression ~\eqref{eq:drho_vacancy} is a special case of Eq.~\eqref{eq:drho}; the corresponding real-space pattern is shown in Fig.~\ref{fig:analytical}.b and it  is in full agreement with numerics (Fig.~\ref{fig:vacancy}c)  and STM data (Fig.~\ref{fig:STM}a).  Its Fourier transform reveals both fermion doubling (Fig.~\ref{fig:analytical}a) and a single-dislocation interference pattern (Fig.~\ref{fig:analytical}b), giving a winding number $\nu = \pm 1$.  Solving $\mathcal{D}\psi = 0$ confirms $\mathrm{Index}\,\mathcal{D} = \nu = \pm 1$~\cite{Abulafia2025}, corresponding to $N_{\mathrm{zm}} = 1$ via Eq.~\eqref{eq:chain}.  To make the winding-number extraction explicit and quantitative, we numerically evaluate Eq.~\eqref{eq:winding} by integrating $\nabla\chi$ along circular contours of increasing radius that encircle the defect in the filtered tight-binding density.  The results are shown in Fig.~\ref{fig:winding}: the vacancy yields $\nu = 1$ to within numerical precision for every contour radius $r/a \in [2,18]$, directly confirming contour independence.  The adatom, subjected to the identical procedure, yields a winding that oscillates and decays to zero at large $r$ — a direct visual demonstration that the two cases are topologically distinct without requiring a change of protocol.

\subsection{Adatom potential: a non-topological counterexample} 
\begin{figure}[t]
  \centering
  \subfloat[]{\includegraphics[width=0.515\columnwidth]{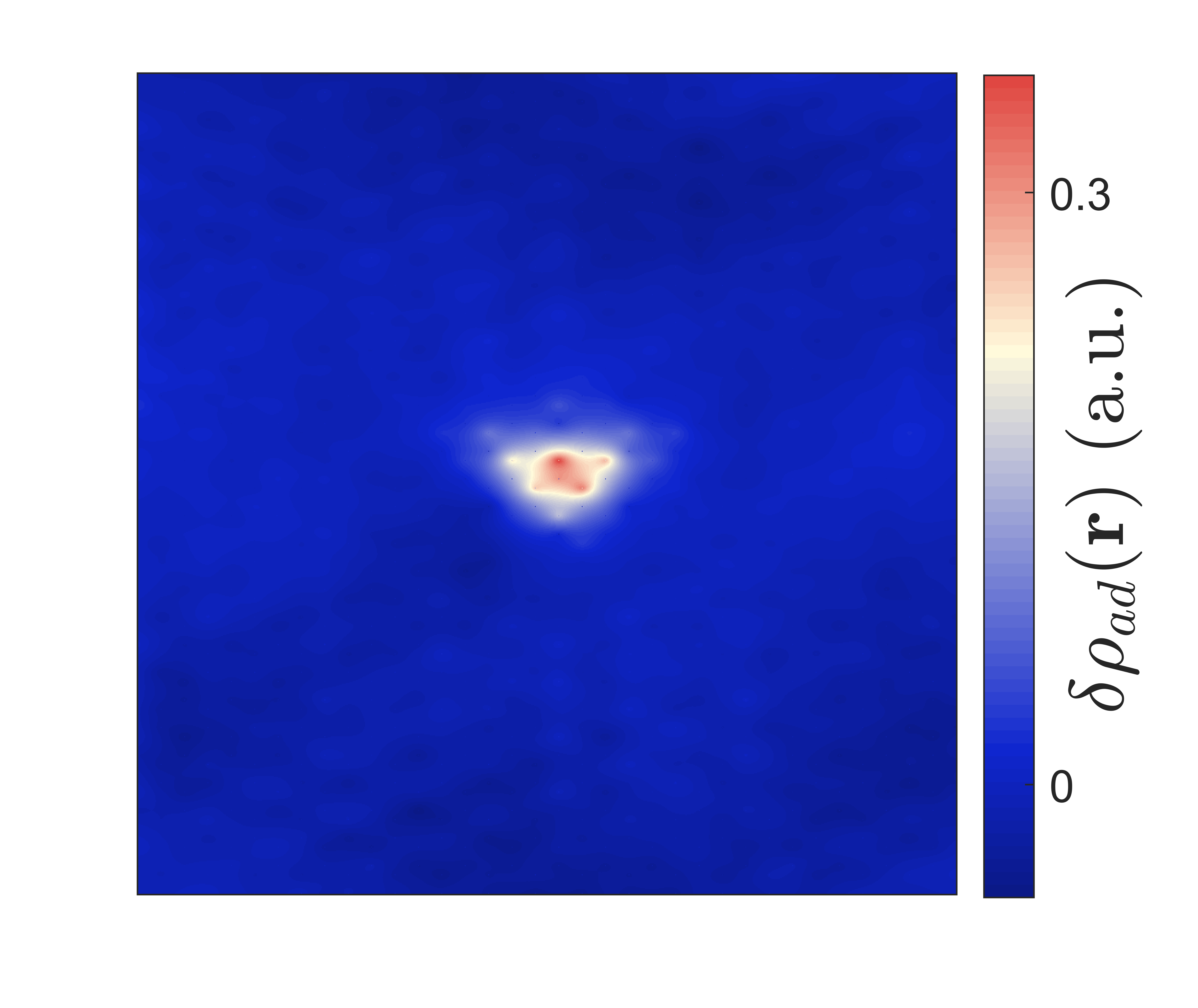}}
  \hfill
  \subfloat[]{\includegraphics[width=0.48\columnwidth]{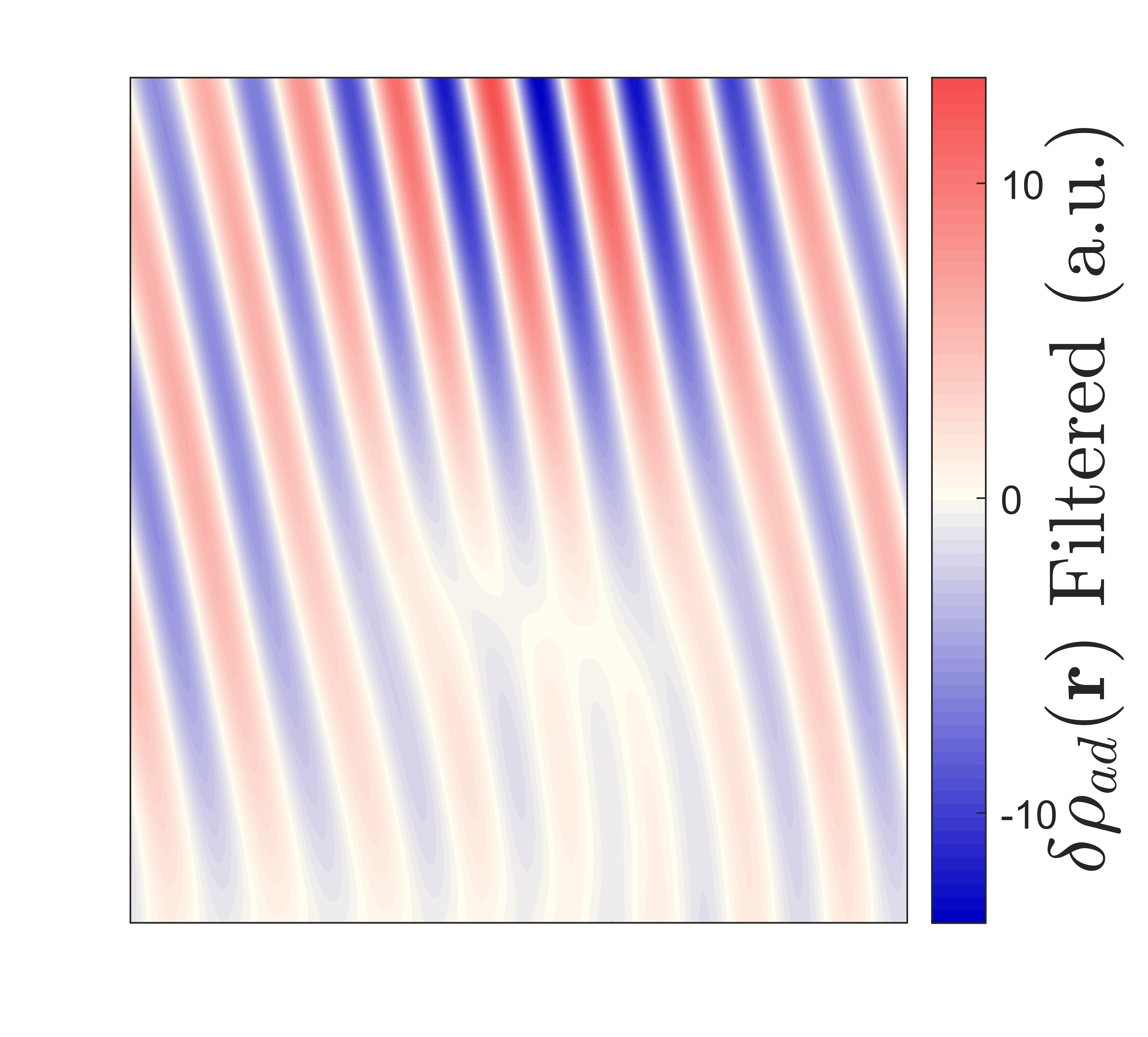}}
  \caption{Graphene with an adatom ($\varepsilon_F = 0.6t$, $V_0 = -3t$).
  (a)~Local electronic density from tight-binding simulation of a $36\times56$ lattice.
  (b)~Filtered image after subtracting the pristine density $\rho_0$ and focusing on two Dirac points; no wavefront dislocation is present.}
  \label{fig:adatom}
\end{figure}

To demonstrate the discriminating power of this method, we examine an adatom potential that breaks chiral symmetry and is not expected to harbor TQS~\cite{Abulafia2025}.  STM images and tight-binding calculations (Fig.~\ref{fig:adatom}a) appear superficially similar to the vacancy case.  However, applying the same filtering yields a markedly different result (Fig.~\ref{fig:adatom}b): the interference pattern shows \emph{no} wavefront dislocation.

\begin{figure}[t]
  \centering
  \subfloat[]{\includegraphics[width=0.52\columnwidth]{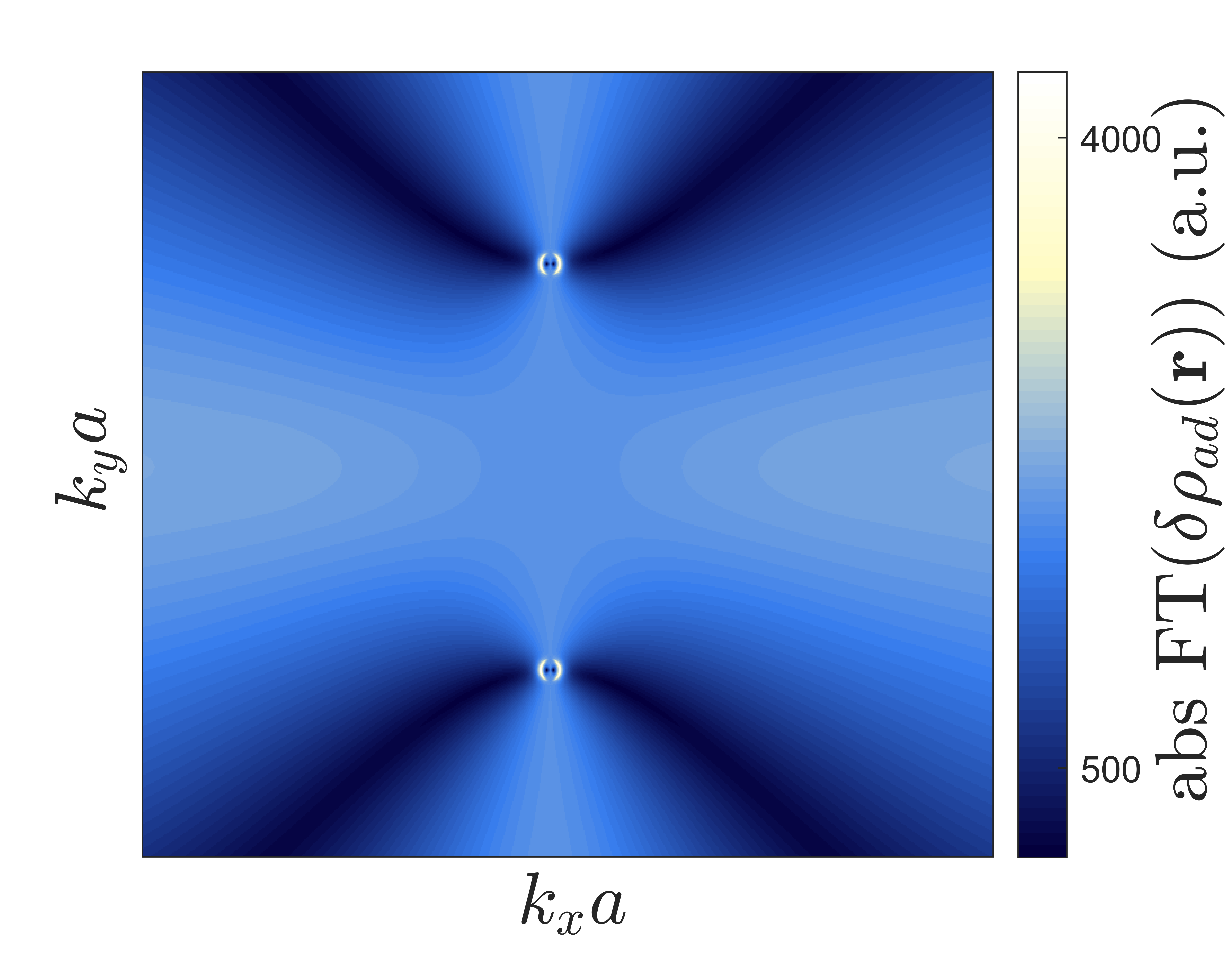}}
  \hfill
  \subfloat[]{\includegraphics[width=0.48\columnwidth]{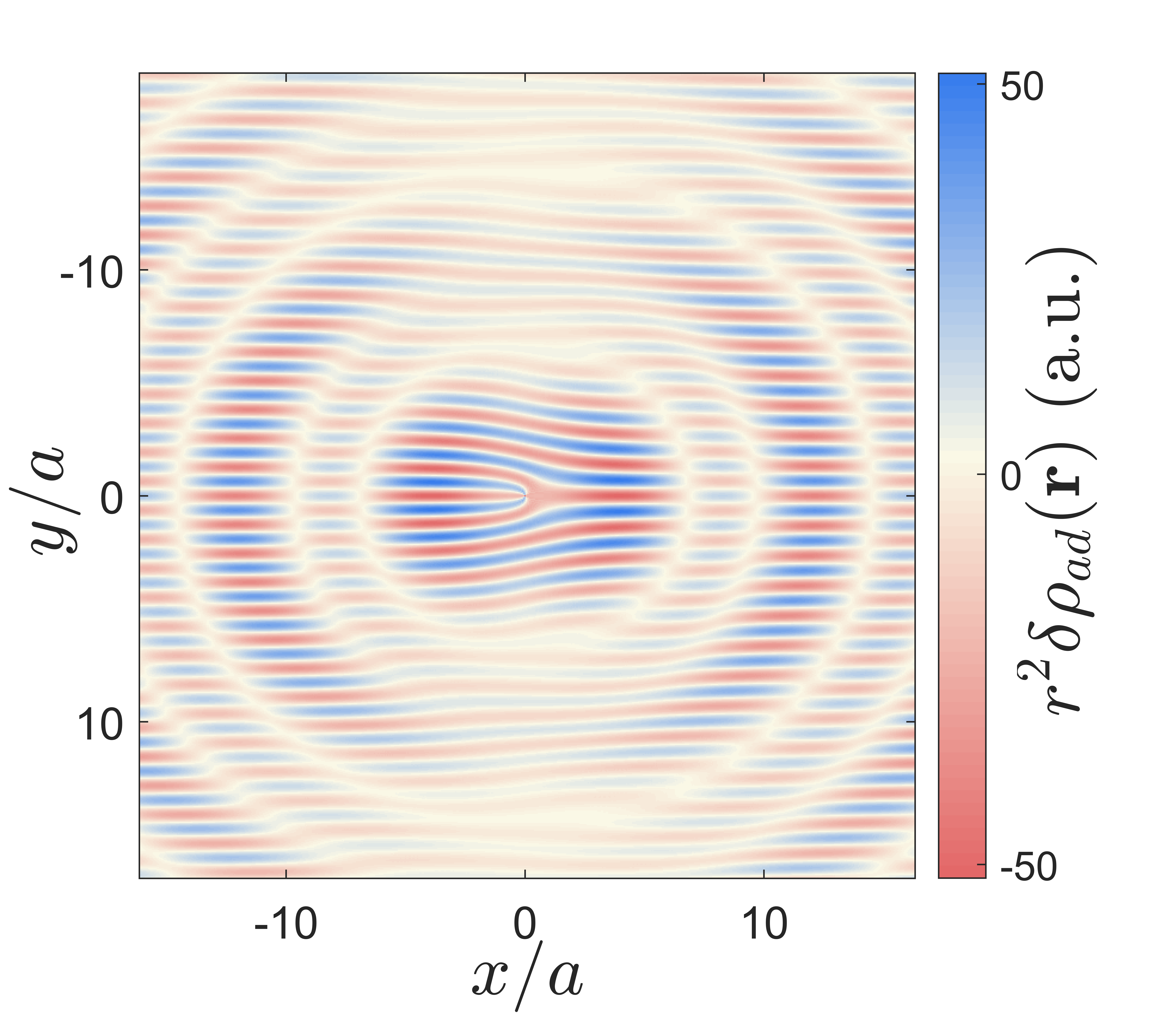}}\\[4pt]
  \subfloat[]{\includegraphics[width=0.48\columnwidth]{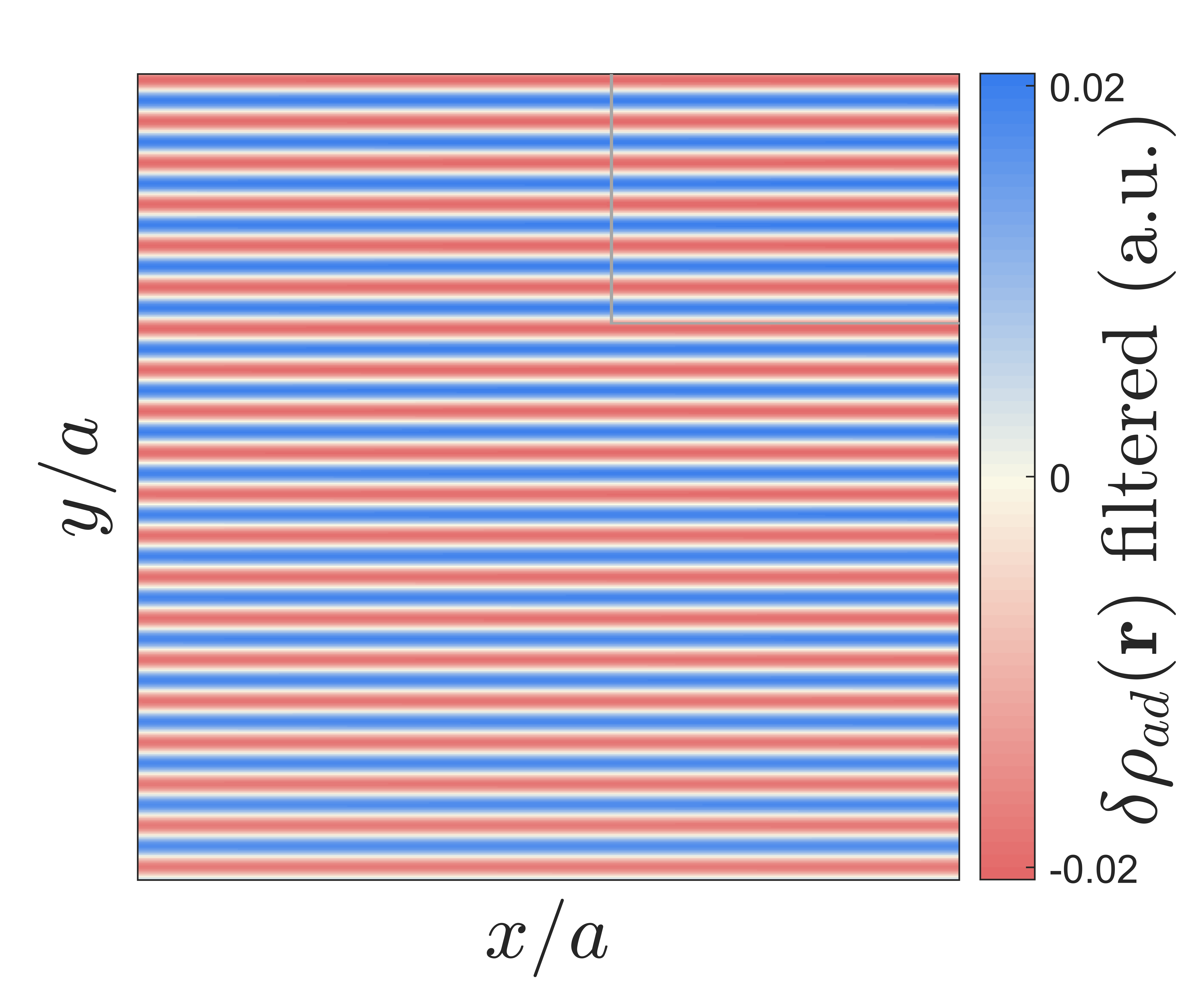}}
  \caption{Adatom density from Eq.~\eqref{eq:drho_adatom}.
  (a)~$|\mathrm{FT}(\delta\rho_{\mathrm{ad}})|$ showing Friedel oscillation rings around the Dirac points in addition to the intervalley peaks.
  (b)~$r^2\delta\rho_{\mathrm{ad}}(\bm{r})$ with $\varepsilon_F = 0.6t$, $V_0 = -3t$; the $r^2$ prefactor is used here purely for visualization to compensate the faster amplitude decay, while the filtering protocol uses $r\delta\rho$ in both cases.
  (c)~After applying the same two-peak filter as for the vacancy: no dislocations are detected.}
  \label{fig:adatom_detail}
\end{figure}

The intervalley density variation for the adatom is~\cite{Abulafia2025,Dutreix2019}
\begin{equation}
\delta\rho_{\mathrm{ad}}(\bm{r}) = f_A(k_F r)\cos\chi_0(\bm{r}) - f_B(k_F r)\cos\chi_2(\bm{r}),
\label{eq:drho_adatom}
\end{equation}
where the radial functions $f_{A,B}$ have similar long-range behavior to $F(r)$ in Eq.~\eqref{eq:drho_vacancy}.  Applying the same protocol, the Fourier transform (Fig.~\ref{fig:adatom_detail}a) shows the expected coupling between the Dirac points, but now surrounded by additional rings arising from Friedel oscillations~\cite{Dutreix2019} produced by the broken chiral symmetry.  These rings are not intervalley-scattering satellites: they encode intravalley charge redistribution at wavevector $2k_F$.  The two-peak filter window therefore automatically excludes them, and the resulting filtered image (Fig.~\ref{fig:adatom_detail}c) shows no dislocation, confirming the absence of TQS without requiring any modification of the protocol.

\begin{figure}[t]
  \centering
  \includegraphics[width=\columnwidth]{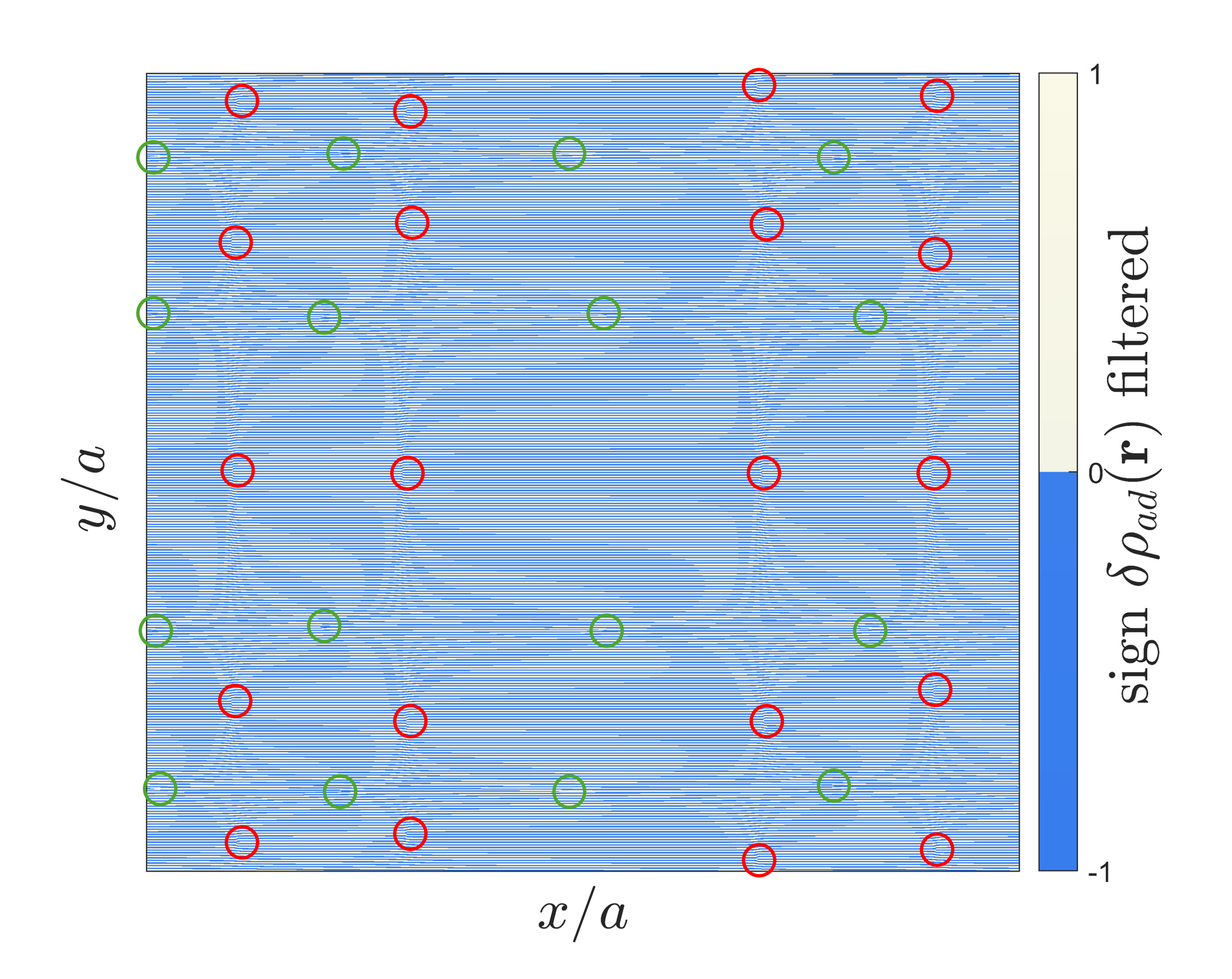}
  \caption{Large-distance behavior of the adatom case after filtering out Friedel oscillations. Dislocations of alternating sign ($\pm 1$, green and red circles) average to zero, confirming the absence of a global topological invariant.}
  \label{fig:friedel}
\end{figure}

It is instructive to contrast the two cases.  For the vacancy, the dislocation constitutes a wavefunction singularity extending to the system boundaries, invariant under chirality-preserving perturbations.  The winding number~$\nu$ is therefore contour-independent.  For the adatom, local dislocations may arise but are confined by Friedel oscillations, preventing a contour-independent winding number.  At large distances (Fig.~\ref{fig:friedel}), dislocations with alternating $\pm 1$ windings average to zero for infinite systems.  We emphasize the distinction between the phase $\chi(\bm{r},n)$ and its winding, compared to the geometric Berry phase around Dirac points in pristine graphene~\cite{Dutreix2019}: only the former reveals the topology of the full Hamiltonian.  
This contrast is also relevant to charge-density dislocations observed near defects in non-topological scenarios~\cite{Zhang2020,Zhang2021}, where Friedel oscillations confine the dislocation pattern to a finite region around the defect, preventing the phase winding from persisting to large distances and thus precluding any contour-independent topological invariant~\cite{Dutreix2019,Engstrom2025}.

\section{Discussion}
We have presented a general method to identify and measure topological invariants through wavefront dislocations in the local electronic density. The connection between ASIT and STM imaging provides a direct, quantitative route to determine winding numbers without relying on indirect transport or spectral signatures. The method distinguishes topological from non-topological states by the presence or absence of contour-independent dislocations in~$\delta\rho(\bm{r})$, demonstrated here through a reanalysis of existing graphene STM data~\cite{Ugeda2010} and validated against tight-binding numerics.  This reanalysis constitutes a proof of concept: the raw STM images already contain the topological information, recoverable through the filtering procedure described here without any new instrumentation.  This result establishes the condensed matter foundation for a broader class of topological diagnostics based on wavefront dislocations.

The mechanism identified here, that the dislocation in any observable built from zero modes inherits a contour-independent phase winding equal to $\nu$, extends beyond the local electronic density.  Since $\delta\rho(\bm{r})$ is determined by the zero-mode wavefunctions via Eq.~\eqref{eq:drho_zeromodes}, the same filtering and winding-number protocol applies equally to the wavefunction~$\psi(\bm{r})$ itself in platforms where it is directly measurable, such as AMO systems~\cite{Faktor2026}.  The principle also extends beyond chiral symmetry classes: in Chern insulators, a vacancy induces a $\mathbb{Z}_2$ defect invariant $\nu = C\cdot m \bmod 2$, and wavefunction dislocations signal this binary charge in direct analogy with the present results~\cite{Makwana2026}.  More broadly, the same index-theorem structure, zero modes, dislocation, contour-independent winding, has been identified in the geometric phase $\gamma_{2q}(\alpha_0,\beta_0)$ of two-qubit systems, where Bell states at Weyl chamber phase boundaries are the zero-mode analogs and dislocations on the Schmidt sphere directly measure the topological invariant of the entangling gate~\cite{Orion2026}.  In each case the condensed matter result established here provides both the conceptual template and the mathematical framework.

Within condensed matter, the next natural arena is Bogoliubov-de Gennes superconductors, where the chiral operator~$\mathcal{D}$ acts on Nambu spinors and the winding number counts Majorana zero modes; the same STM protocol applied to the local density of states would directly measure the topological invariant of the superconducting phase, identifying Majorana-hosting vortices through their characteristic dislocation pattern.  More complex scenarios, including higher winding numbers, will be explored in subsequent work.

\begin{acknowledgments}
This research was funded by the Israel Science Foundation Grant No.~772/21 and the Pazy Foundation.
\end{acknowledgments}
\appendix
\section{Derivation of the zero-mode dominance equation}\label{appendix:zero_mode_eq}

Equation~(\ref{eq:drho_zeromodes}) states that the change in local electronic density~$\delta\rho(\bm{r})$ is entirely determined by the zero-energy modes of the chiral Hamiltonian.  We derive it here starting from Eq.~(\ref{eq:rho}),
\begin{equation}
\rho(\bm{r}) = -\frac{1}{\pi}\int_{-\infty}^{\infty} dE\,\mathrm{Im}\sum_n
\frac{|\varphi_n(\bm{r})|^2}{E - E_n + i0^+}.
\label{eq:SM_rho}
\end{equation}
Evaluating the integral using $\mathrm{Im}(E - E_n + i0^+)^{-1} = -\pi\delta(E - E_n)$, the density at chemical potential $\mu$ (measuring occupied states, $E_n \le \mu$) is
\begin{equation}
\rho(\bm{r}) = \sum_{E_n \le \mu} |\varphi_n(\bm{r})|^2.
\end{equation}
For a chiral Hamiltonian~$H = \bigl(\begin{smallmatrix}0&\mathcal{D}^\dagger\\\mathcal{D}&0\end{smallmatrix}\bigr)$, the spectrum has particle-hole symmetry: every nonzero eigenvalue~$E_n > 0$ is accompanied by a partner eigenvalue~$-E_n$, with eigenstates related by $|\varphi_{-E_n}\rangle = \sigma_3|\varphi_{E_n}\rangle$, where $\sigma_3$ is the chiral symmetry operator.  The spatial densities of a pair satisfy $|\varphi_{-E_n}(\bm{r})|^2 = |\varphi_{E_n}(\bm{r})|^2$.

In the absence of the defect potential~$V$, the pristine system has density~$\rho_0(\bm{r})$.  The change $\delta\rho(\bm{r}) = \rho(\bm{r}) - \rho_0(\bm{r})$ receives contributions from all states shifted by~$V$.  Setting the chemical potential at the charge neutrality point ($\mu = 0^-$), the contribution of each $\pm E_n$ pair to $\delta\rho$ is
\begin{equation}
\delta\rho_{\pm E_n}(\bm{r}) = \Delta|\varphi_{-E_n}(\bm{r})|^2 - \Delta|\varphi_{+E_n}(\bm{r})|^2,
\end{equation}
where $\Delta|\varphi_{\pm E_n}|^2$ denotes the change in the squared wavefunction amplitude due to the defect.  Because $|\varphi_{-E_n}|^2 = |\varphi_{E_n}|^2$ is enforced by chiral symmetry at every point~$\bm{r}$, these two contributions cancel exactly:
\begin{equation}
\delta\rho_{\pm E_n}(\bm{r}) = 0 \quad \text{for every } E_n \neq 0.
\end{equation}
The only surviving contributions are from states at exactly $E_n = 0$, the zero modes, which are \emph{not} subject to the $\pm E_n$ pairing constraint.  Summing over all zero modes (labeled by index~$n$) with their sublattice sign,
\begin{equation}
\delta\rho(\bm{r}) = \frac{e}{2}\sum_n \mathrm{sign}(E_n)\,|\varphi_n(\bm{r})|^2,
\label{eq:SM_drho_zm}
\end{equation}
where the factor $e/2$ arises from the charge neutrality convention (each zero mode contributes half an electron charge), and $\mathrm{sign}(E_n)$ distinguishes zero modes on the two sublattices (those lifted to $+m$ vs $-m$ by a mass term $m\sigma_3$).  This is Eq.~(\ref{eq:drho_zeromodes}).

In summary, the dislocation pattern in $\delta\rho(\bm{r})$ is a direct imprint of the topological zero-mode wavefunctions $\varphi_n(\bm{r})$.  STM, which measures $\delta\rho(\bm{r})$, thereby provides access to the zero-mode phase structure without needing to resolve individual wavefunctions.

\section{Tight-binding numerics}

We used a nearest-neighbor tight-binding model on the honeycomb lattice.  For a lattice of~$N$ sites, the Hamiltonian is an $N\times N$ matrix with hopping elements~$t$ between each site and its three neighbors.  When additional energy scales appear (e.g., an adatom), all parameters are expressed in units of~$t$, with the adatom on-site energy set to $V_0 = -3t$.  Periodic boundary conditions are imposed to suppress supernumerary zero modes.  A vacancy is implemented by deleting all hoppings to the missing site, producing a row and column of zeros; removing these yields an $(N-1)\times(N-1)$ Hamiltonian.

\section{Local density calculation}

After constructing the Hamiltonian matrix, we obtain eigenvalues and eigenstates in the lattice-site basis.  The local electronic density is computed from the imaginary part of the Green's function (Eq.~(\ref{eq:rho})), where the energy integral is discretized with $dE = 10^{-4}t$ over the symmetric range $[-t,t]$ to preserve particle-hole symmetry.  Since the zero mode lies at $\pm 10^{-16}t$, symmetric bounds ensure its inclusion.  By chiral symmetry $\{H,\sigma_3\}=0$, contributions from positive and negative energies are equal, justifying the range extension.  Numerically, we set $i0^+ = 10^{-5}t$.

\section{Orbital reconstruction from tight binding}

Our simulations use the nearest-neighbor tight-binding approximation, where each atom is a single lattice site without orbital structure.  Extracting wavefronts and their dislocations as observed in STM requires a continuous description.  We recover orbital information as follows.  The tight-binding model produces discrete wavefunction amplitudes~$\phi_j$ at each lattice site~$j$.  To approximate continuous space, we embed the lattice in a finer grid, with each lattice site contributing an exponentially decaying weight to surrounding pixels,
\begin{equation}
w(\bm{r}) \propto \exp\!\left(-\frac{|\bm{r} - \bm{R}_j|}{a}\right),
\end{equation}
where $a$ is the lattice constant (the decay length is identical for both the vacancy and adatom cases).  The resulting smooth density field is then cropped around the defect of interest.  This approach enables efficient simulation of large systems (e.g., 2052 sites).

\section{Filtering protocol: detailed steps and reproducibility notes}

The filtering protocol extracts the topological dislocation signal from $\delta\rho(\bm{r})$ in four steps, described here with sufficient detail for independent reproduction.

\textbf{Step~1: $r$-weighting.}  Starting from the continuum data for $\delta\rho(\bm{r})$ on a uniform real-space grid, we multiply pointwise by the distance~$r$ from the defect, forming $g(\bm{r}) \equiv r\,\delta\rho(\bm{r})$.  Since the relevant Friedel oscillations decay as~$r^{-1}$ for 2D Dirac fermions, the product~$g(\bm{r})$ tends to a spatially oscillating function of approximately constant amplitude at large~$r$, concentrating spectral weight into sharp peaks rather than diffuse rings.  For visualization purposes only, Fig.~\ref{fig:adatom}b displays $r^2\delta\rho_{\mathrm{ad}}(\bm{r})$ to compensate the faster amplitude decay of the adatom density; however, the filtering in Step~2 is performed on $r\delta\rho$ in both the vacancy and adatom cases.

\textbf{Step~2: Fourier transform.}  We compute the discrete 2D FFT of~$g(\bm{r})$ on the numerical grid using a standard FFT algorithm (MATLAB \texttt{fft2}).  We record the full \emph{complex} spectrum $\hat{g}(\bm{k}) = |\hat{g}(\bm{k})|\,e^{i\phi(\bm{k})}$, retaining both amplitude and phase.  The magnitude $|\hat{g}(\bm{k})|$ is shown in Fig.~\ref{fig:vacancy}b.  The phase $\phi(\bm{k})$ is not displayed but is essential for Step~4.

\textbf{Step~3: Peak selection and masking.}  We identify the two diametrically opposite intervalley satellite peaks at $+\Delta\bm{K}$ and $-\Delta\bm{K}$ in $|\hat{g}(\bm{k})|$ by locating the two highest-amplitude off-origin maxima related by inversion symmetry $\bm{k}\to-\bm{k}$ (a consequence of $\delta\rho$ being real).  We construct a binary mask equal to~1 within a square window of $8\times8$ pixels centered on each selected peak, and~0 everywhere else, including the dc component at $\bm{k}=0$.  The window size of~8 pixels was chosen to encompass the full peak width while excluding neighboring structure; results are insensitive to this choice over the range~6--12 pixels.  We apply the mask to the full complex spectrum: $\hat{g}_{\mathrm{filt}}(\bm{k}) = M(\bm{k})\,\hat{g}(\bm{k})$.  For the adatom case, the Friedel oscillation rings visible in Fig.~\ref{fig:adatom_detail}a lie at wavevector $2k_F \neq \Delta K$ and are therefore automatically excluded by this window without any modification of the protocol.

\textbf{Step~4: Inverse Fourier transform and phase extraction.}  We compute the inverse FFT of the complex filtered spectrum $\hat{g}_{\mathrm{filt}}(\bm{k})$, yielding the complex real-space field $g_{\mathrm{filt}}(\bm{r}) = \mathrm{IFT}[\hat{g}_{\mathrm{filt}}](\bm{r})$.  It is critical that the \emph{complex} spectrum is used: discarding the phase $\phi(\bm{k})$ before inverting would suppress the dislocation, since the wavefront singularity is encoded in the phase variation of~$\hat{g}$ near $\pm\Delta\bm{K}$.  Physically, the inverse transform isolates the two-valley contribution to~$\delta\rho$, and the real part of~$g_{\mathrm{filt}}(\bm{r})$ is the filtered density shown in Figs.~\ref{fig:vacancy}c, \ref{fig:analytical}b, \ref{fig:STM}b, \ref{fig:adatom}b, and~\ref{fig:adatom_detail}b.

\textbf{Phase extraction for the winding-number contour integral.}  To evaluate Eq.~(\ref{eq:winding}) numerically, we extract the phase field $\chi(\bm{r}) = \arg[g_{\mathrm{filt}}(\bm{r})]$ from the complex filtered image.  Near the dislocation core the argument function has a branch cut, and naive differentiation of $\arg[g_{\mathrm{filt}}]$ introduces $\pm 2\pi$ jumps.  We handle this using standard 2D phase unwrapping (Flynn's minimum discontinuity algorithm~\cite{Flynn1997}), which removes the branch cuts by adding multiples of $2\pi$ to enforce continuity. The gradient $\nabla\chi$ is then computed by finite differences on the unwrapped phase map.  The contour integral $\frac{1}{2\pi}\oint_\mathcal{C}\nabla\chi\cdot d\bm{r}$ is evaluated numerically along a discrete circular path of radius~$r$ by summing the gradient components along the path and multiplying by the pixel spacing.  This procedure is repeated for contour radii $r/a \in [2,18]$ for both the vacancy and adatom cases; the results are shown in Fig.~\ref{fig:winding}.

\textbf{STM data preprocessing.}  The STM data of Ref.~\cite{Ugeda2010} were obtained from the published figures.  Before applying the filtering protocol, a standard plane subtraction was performed to remove the slow linear background gradient arising from STM tip drift.  No smoothing was applied.  The same Steps~1--4 were then followed as for the numerical data, with the only difference that the pristine density $\rho_0$ is not separately available from experiment; instead, we treat the raw STM image as $\delta\rho$ directly, which is valid since the vacancy-induced signal dominates the charge-neutrality-point image.


\begin{figure*}[t]
\centering
\subfloat[]{\includegraphics[width=0.1\textwidth,height=4cm]{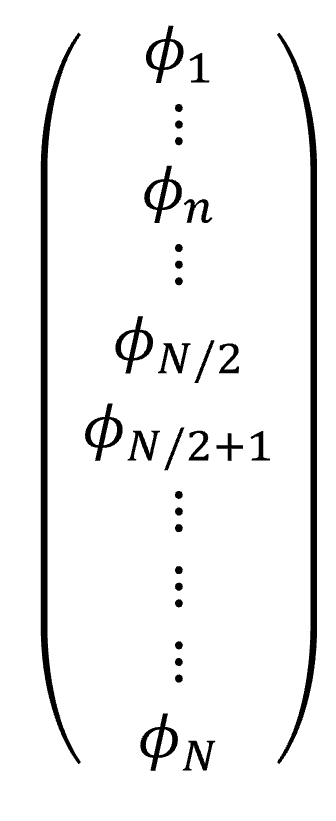}}\:\:
\subfloat[]{\includegraphics[width=0.4\textwidth,height=4cm]{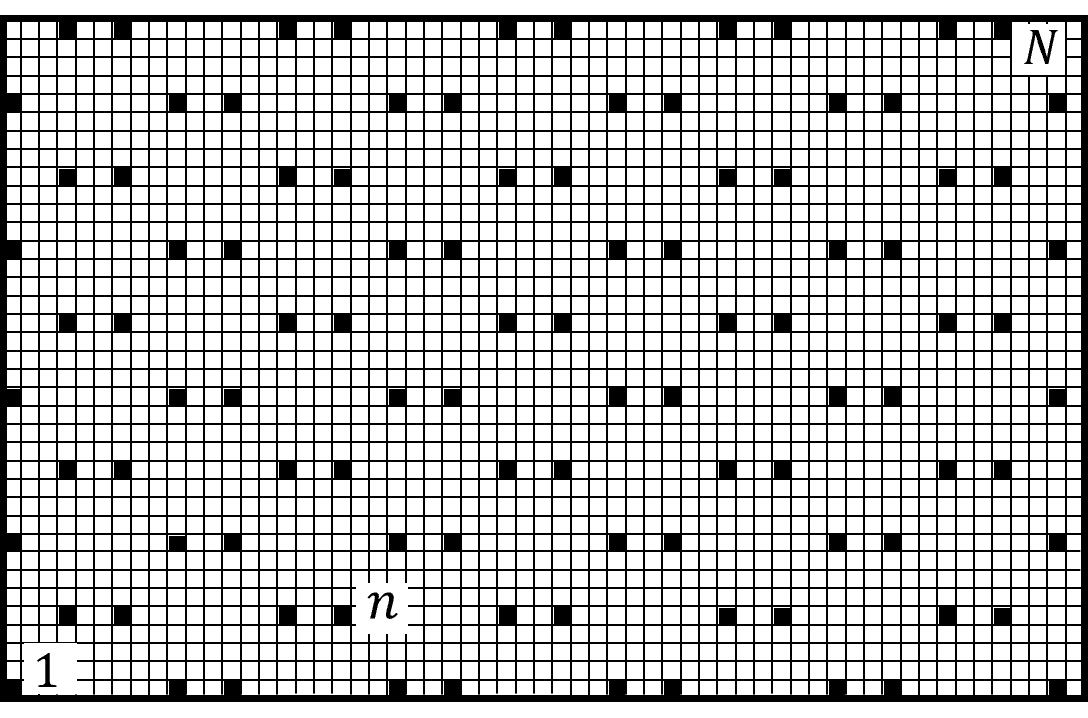}}\:\:
\subfloat[]{\includegraphics[width=0.4\textwidth,height=4cm]{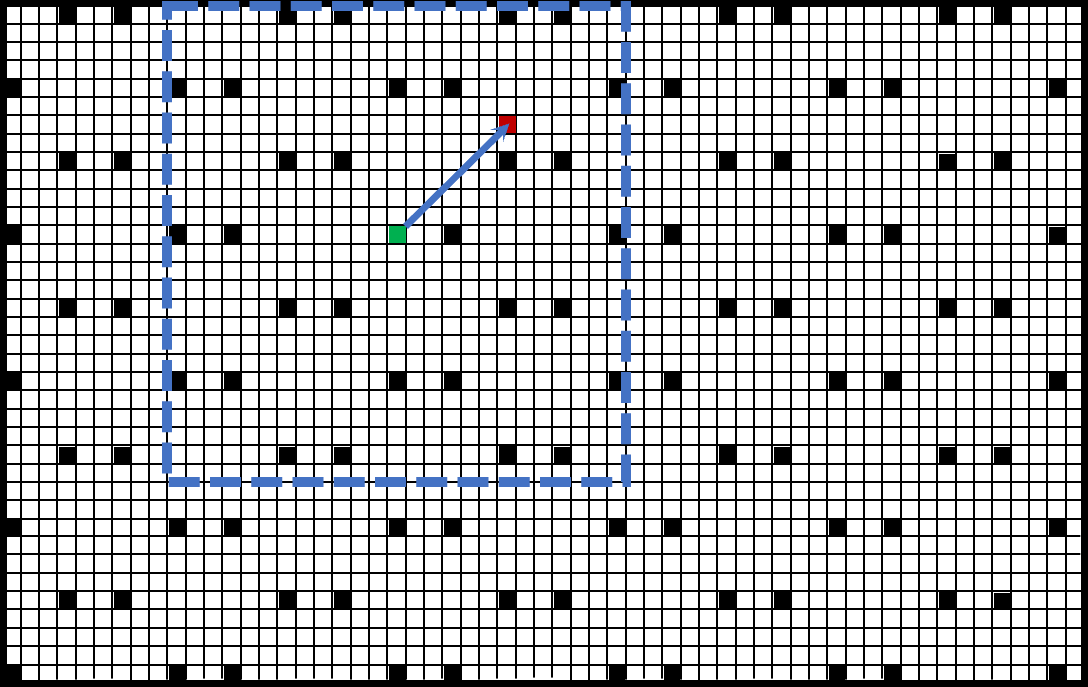}}\\
\subfloat[]{\includegraphics[width=0.85\textwidth,height=5cm]{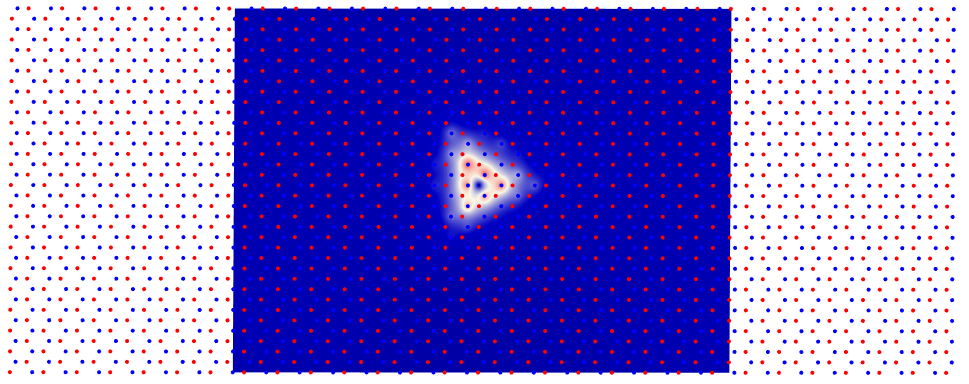}}
\caption{Orbital reconstruction from tight-binding simulation.
(a)~Discrete wavefunction vector; entries correspond to sublattices $A$ and $B$.
(b)~High-resolution pixel grid; black squares are original lattice sites.
(c)~Exponentially decaying orbital weight centered on one site (green); contribution to the red pixel decays as $e^{-|\bm{r}-\bm{R}_j|/a}$ with decay length equal to the lattice constant~$a$.
(d)~Full reconstructed density image, cropped around the vacancy.}
\label{fig:sparial_behavior_supp}
\end{figure*}

\bibliography{Nature.bib}

@article{Hashimoto2008,
  title = {Quantum Hall Transition in Real Space: From Localized to Extended States},
  author = {Hashimoto, K. and Sohrmann, C. and Wiebe, J. and Inaoka, T. and Meier, F. and Hirayama, Y. and R\"omer, R. A. and Wiesendanger, R. and Morgenstern, M.},
  journal = {Phys. Rev. Lett.},
  volume = {101},
  issue = {25},
  pages = {256802},
  numpages = {4},
  year = {2008},
  month = {Dec},
  publisher = {American Physical Society},
  doi={http://doi.org/10.1103/PhysRevLett.101.256802},
  url = {https://link.aps.org/doi/10.1103/PhysRevLett.101.256802},
howpublished = {\url{https://link.aps.org/doi/10.1103/PhysRevLett.101.256802}},
}

@article{Ugeda2010,
  title = {Missing Atom as a Source of Carbon Magnetism},
  author = {Ugeda, M. M. and Brihuega, I. and Guinea, F. and G\'omez-Rodr\'iguez, J. M.},
  journal = {Phys. Rev. Lett.},
  volume = {104},
  issue = {9},
  pages = {096804},
  numpages = {4},
  year = {2010},
  month = Mar,
  publisher = {American Physical Society},
  doi = {https://doi.org/10.1103/PhysRevLett.104.096804},
  url = {https://link.aps.org/doi/10.1103/PhysRevLett.104.096804}
}

@article{Ovdat2020,
  title = {Vacancies in graphene: {Dirac} physics and fractional vacuum charges},
  author = {Ovdat, Omrie and Don, Yaroslav and Akkermans, Eric},
  journal = {Phys. Rev. B},
  volume = {102},
  issue = {7},
  pages = {075109},
  numpages = {6},
  year = {2020},
  month = Aug,
  publisher = {American Physical Society},
  doi = {https://doi.org/10.1103/PhysRevB.102.075109},
  url = {https://link.aps.org/doi/10.1103/PhysRevB.102.075109}
}

@article{Dutreix2019, title = {Measuring the Berry phase of graphene from wavefront dislocations in Friedel oscillations}, author = {Dutreix, C. and GonzÃ¡lez-Herrero H. and Brihuega, I. and Katsnelson, M. I. and Chapelier, C. and Renard, V. T. }, journal = {Nature}, volume = {574}, issue = {7777}, pages = {219-222}, numpages = {}, year = {2019}, month = {Oct}, doi={http://doi.org/10.1038/s41586-019-1613-5}, url = {https://doi.org/10.1038/s41586-019-1613-5} }

@article{Zhang2020, title = {Local Berry Phase Signatures of Bilayer Graphene in Intervalley Quantum Interference}, author = {Zhang, Yu and Su, Ying and He, Lin}, journal = {Phys. Rev. Lett.}, volume = {125}, issue = {11}, pages = {116804}, numpages = {6}, year = {2020}, month = {Sep}, publisher = {American Physical Society}, doi={http://doi.org/10.1103/PhysRevLett.125.116804}, url = {https://link.aps.org/doi/10.1103/PhysRevLett.125.116804}
}

@article{Atiyah1963,
author = {M. F. Atiyah and I. M. Singer},
title = {{The index of elliptic operators on compact manifolds}},
volume = {69},
journal = {Bulletin of the American Mathematical Society},
number = {3},
publisher = {American Mathematical Society},
pages = {422 -- 433},
year = {1963},
URL={https://www.ams.org/journals/bull/1963-69-03/S0002-9904-1963-10957-X/},
 doi={http://ams.org/journals/bull/1963-69-03/S0002-9904-1963-10957-X/},
}

@article{Teo2010,
  title = {Topological defects and gapless modes in insulators and superconductors},
  author = {Teo, Jeffrey C. Y. and Kane, C. L.},
  journal = {Phys. Rev. B},
  volume = {82},
  issue = {11},
  pages = {115120},
  numpages = {26},
  year = {2010},
  month = {Sep},
  publisher = {American Physical Society},
  doi={http://doi.org/10.1103/PhysRevB.82.115120},
  url = {https://link.aps.org/doi/10.1103/PhysRevB.82.115120}
}

@article{Altland1997,
author = {Altland, Alexander and Zirnbauer, Martin R.},
doi={http://doi.org/10.1103/PhysRevB.55.1142},
issn = {0163-1829},
journal = {Physical Review B},
month = {jan},
number = {2},
pages = {1142--1161},
title = {{Nonstandard symmetry classes in mesoscopic normal-superconducting hybrid structures}},
url = {https://link.aps.org/doi/10.1103/PhysRevB.55.1142},
volume = {55},
year = {1997}
}

@article{Chiu2016,
  title = {Classification of topological quantum matter with symmetries},
  author = {Chiu, Ching-Kai and Teo, Jeffrey C. Y. and Schnyder, Andreas P. and Ryu, Shinsei},
  journal = {Rev. Mod. Phys.},
  volume = {88},
  issue = {3},
  pages = {035005},
  numpages = {63},
  year = {2016},
  month = {Aug},
  publisher = {American Physical Society},
  doi={http://doi.org/10.1103/RevModPhys.88.035005},
  url = {https://link.aps.org/doi/10.1103/RevModPhys.88.035005}
}

@article{Kitaev2009,
author = {Kitaev,Alexei },
title = {Periodic table for topological insulators and superconductors},
journal = {AIP Conference Proceedings},
volume = {1134},
number = {1},
pages = {22-30},
year = {2009},
doi={http://doi.org/10.1063/1.3149495},
URL = { 
        https://aip.scitation.org/doi/abs/10.1063/1.3149495 
}
}

@article{KOHMOTO1985,
title = {Topological invariant and the quantization of the Hall conductance},
journal = {Annals of Physics},
volume = {160},
number = {2},
pages = {343-354},
year = {1985},
issn = {0003-4916},
doi={https://doi.org/10.1016/0003-4916(85)90148-4},
url = {https://www.sciencedirect.com/science/article/pii/0003491685901484},
author = {Mahito Kohmoto}
}

@article{Schnyder2008,
  title = {Classification of topological insulators and superconductors in three spatial dimensions},
  author = {Schnyder, Andreas P. and Ryu, Shinsei and Furusaki, Akira and Ludwig, Andreas W. W.},
  journal = {Phys. Rev. B},
  volume = {78},
  issue = {19},
  pages = {195125},
  numpages = {22},
  year = {2008},
  month = {Nov},
  publisher = {American Physical Society},
  doi={http://doi.org/10.1103/PhysRevB.78.195125},
  url = {https://link.aps.org/doi/10.1103/PhysRevB.78.195125}
}

@article{Goft2023,
  title = {Defects in graphene: A topological description},
  author = {Goft, Amit and Abulafia, Yuval and Orion, Nadav and Schochet, Claude L. and Akkermans, Eric},
  journal = {Phys. Rev. B},
  volume = {108},
  issue = {5},
  pages = {054101},
  numpages = {10},
  year = {2023},
  month = {Aug},
  publisher = {American Physical Society},
  doi={http://doi.org/10.1103/PhysRevB.108.054101},
  url = {https://link.aps.org/doi/10.1103/PhysRevB.108.054101}
}

@article{Atiyah1968,
 ISSN = {0003486X},
 url = {http://www.jstor.org/stable/1970715},
 author = {M. F. Atiyah and I. M. Singer},
 journal = {Annals of Mathematics},
 number = {3},
 pages = {484--530},
 publisher = {Annals of Mathematics},
 doi={http://jstor.org/stable/1970715},
 title = {The Index of Elliptic Operators: I},
 urldate = {2023-04-02},
 volume = {87},
 year = {1968}
}

@book{Nakahara1990,
      author        = "Nakahara, Mikio",
      title         = "{Geometry, topology and physics}",
      publisher     = "Hilger",
      address       = "Bristol",
      series        = "Graduate student series in physics",
      year          = "1990",
      url           = "https://cds.cern.ch/record/206619",
        doi= "http://cds.cern.ch/record/206619",
}

@book{yankowsky2013,
  title={Quantum Field Theory and Topology},
  author={Yankowsky, E. and Schwarz, A.S. and Levy, S.},
  isbn={9783662029435},
  series={Grundlehren der mathematischen Wissenschaften},
  url={https://books.google.co.il/books?id=RhXpCAAAQBAJ},
  year={2013},
  publisher={Springer Berlin Heidelberg}
}

@article{Dutreix2017, title = {Geometrical phase shift in Friedel oscillations}, author = {Dutreix, C. and Delplace, P.}, journal = {Phys. Rev. B}, volume = {96}, issue = {19}, pages = {195207}, numpages = {10}, year = {2017}, month = {Nov}, publisher = {American Physical Society}, doi={http://doi.org/10.1103/PhysRevB.96.195207}, url = {https://link.aps.org/doi/10.1103/PhysRevB.96.195207} }

@article{Dutreix2021,
  title = {Wavefront dislocations reveal the topology of quasi-1D photonic insulators},
  author = {Dutreix, ClÃ©ment and Bellec Matthieu and Delplace Pierre and Mortessagne Fabrice},
  journal = {Nature Communications},
  volume = {12},
  issue = { 1},
  pages = {3571},
  year = {2021},
  month = {Jun},
  publisher = {American Physical Society},
  doi={http://doi.org/10.1038/s41467-021-23790-w},
  url = {https://doi.org/10.1038/s41467-021-23790-w
}
}

@article{Qi2011,
  title = {Topological insulators and superconductors},
  author = {Qi, Xiao-Liang and Zhang, Shou-Cheng},
  journal = {Rev. Mod. Phys.},
  volume = {83},
  issue = {4},
  pages = {1057--1110},
  numpages = {0},
  year = {2011},
  month = {Oct},
  publisher = {American Physical Society},
  doi={http://doi.org/10.1103/RevModPhys.83.1057},
  url = {https://link.aps.org/doi/10.1103/RevModPhys.83.1057}
}

@article{Hasan2010,
  title = {Colloquium: Topological insulators},
  author = {Hasan, M. Z. and Kane, C. L.},
  journal = {Rev. Mod. Phys.},
  volume = {82},
  issue = {4},
  pages = {3045--3067},
  numpages = {0},
  year = {2010},
  month = {Nov},
  publisher = {American Physical Society},
  doi={http://doi.org/10.1103/RevModPhys.82.3045},
  url = {https://link.aps.org/doi/10.1103/RevModPhys.82.3045}
}

@article{Zhang2021,
  title = {Geometric wavefront dislocations of RKKY interaction in graphene},
  author = {Zhang, Shu-Hui and Yang, Jin and Shao, Ding-Fu and Yang, Wen and Chang, Kai},
  journal = {Phys. Rev. B},
  volume = {104},
  issue = {24},
  pages = {245405},
  numpages = {9},
  year = {2021},
  month = {Dec},
  publisher = {American Physical Society},
  doi={http://doi.org/10.1103/PhysRevB.104.245405},
  url = {https://link.aps.org/doi/10.1103/PhysRevB.104.245405}
}

@article{
Bernevig2006,
author = {B. Andrei Bernevig  and Taylor L. Hughes  and Shou-Cheng Zhang },
title = {Quantum Spin Hall Effect and Topological Phase Transition in HgTe Quantum Wells},
journal = {Science},
volume = {314},
number = {5806},
pages = {1757-1761},
year = {2006},
doi={http://doi.org/10.1126/science.1133734},
URL = {https://www.science.org/doi/abs/10.1126/science.1133734}}

@article{Jackiw2012,
doi={http://doi.org/10.1088/0031-8949/2012/T146/014005},
url = {https://dx.doi.org/10.1088/0031-8949/2012/T146/014005},
year = {2012},
month = {jan},
publisher = {},
volume = {2012},
number = {T146},
pages = {014005},
author = {Jackiw, R},
title = {Fractional and Majorana fermions: the physics of zero-energy modes},
journal = {Physica Scripta}
}

@article{Sarma2015,
doi={http://doi.org/10.1038/npjqi.2015.1},
url = {https://doi.org/10.1038/npjqi.2015.1},
year = {2015},
month = {oct},
publisher = {},
volume = {1},
number = {},
pages = {15001},
author = {Sarma, Sankar Das and Freedman, Michael and Nayak, Chetan},
title = {Majorana zero modes and topological quantum computation},
journal = {npj Quantum Information}
}

@article{Engstrom2025,
  title = {Detecting the topological winding of superconducting nodes via local density of states},
  author = {Engstr\"om, Lena and Simon, Pascal and Mesaros, Andrej},
  journal = {Phys. Rev. B},
  volume = {111},
  issue = {13},
  pages = {134505},
  numpages = {17},
  year = {2025},
  month = {Apr},
  publisher = {American Physical Society},
  doi={http://doi.org/10.1103/PhysRevB.111.134505},
  url = {https://link.aps.org/doi/10.1103/PhysRevB.111.134505}
}

@article{Nielsen1983,
title = {The Adler-Bell-Jackiw anomaly and Weyl fermions in a crystal},
journal = {Physics Letters B},
volume = {130},
number = {6},
pages = {389-396},
year = {1983},
issn = {0370-2693},
doi = {https://doi.org/10.1016/0370-2693(83)91529-0},
url = {https://www.sciencedirect.com/science/article/pii/0370269383915290},
author = {H.B. Nielsen and Masao Ninomiya}
}

@article{Bianco2011,
  title = {Mapping topological order in coordinate space},
  author = {Bianco, Raffaello and Resta, Raffaele},
  journal = {Phys. Rev. B},
  volume = {84},
  issue = {24},
  pages = {241106},
  numpages = {4},
  year = {2011},
  month = {Dec},
  publisher = {American Physical Society},
  doi={http://doi.org/10.1103/PhysRevB.84.241106},
  url = {https://link.aps.org/doi/10.1103/PhysRevB.84.241106}
}

@article{Aidelsburger2015,
	title = {Measuring the {Chern} number of {Hofstadter} bands with ultracold bosonic atoms},
	volume = {11},
	issn = {1745-2481},
	url = {https://doi.org/10.1038/nphys3171},
	doi={http://doi.org/10.1038/nphys3171},
	number = {2},
	journal = {Nature Physics},
	author = {Aidelsburger, M. and Lohse, M. and Schweizer, C. and Atala, M. and Barreiro, J. T. and Nascimbène, S. and Cooper, N. R. and Bloch, I. and Goldman, N.},
	month = feb,
	year = {2015},
	pages = {162--166},
}

@article{
Mancini2015,
author = {M. Mancini  and G. Pagano  and G. Cappellini  and L. Livi  and M. Rider  and J. Catani  and C. Sias  and P. Zoller  and M. Inguscio  and M. Dalmonte  and L. Fallani },
title = {Observation of chiral edge states with neutral fermions in synthetic Hall ribbons},
journal = {Science},
volume = {349},
number = {6255},
pages = {1510-1513},
year = {2015},
doi = {https://doi.org/10.1126/science.aaa8736},
URL = {https://www.science.org/doi/abs/10.1126/science.aaa8736}}

@article{Song2022,
  title = {Magic-Angle Twisted Bilayer Graphene as a Topological Heavy Fermion Problem},
  author = {Song, Zhi-Da and Bernevig, B. Andrei},
  journal = {Phys. Rev. Lett.},
  volume = {129},
  issue = {4},
  pages = {047601},
  numpages = {10},
  year = {2022},
  month = {Jul},
  publisher = {American Physical Society},
  doi = {https://doi.org/10.1103/PhysRevLett.129.047601},
  url = {https://link.aps.org/doi/10.1103/PhysRevLett.129.047601}
}

@article{Calderon2020,
  title = {Interactions in the 8-orbital model for twisted bilayer graphene},
  author = {Calder\'on, M. J. and Bascones, E.},
  journal = {Phys. Rev. B},
  volume = {102},
  issue = {15},
  pages = {155149},
  numpages = {9},
  year = {2020},
  month = {Oct},
  publisher = {American Physical Society},
  doi = {https://doi.org/10.1103/PhysRevB.102.155149},
  url = {https://link.aps.org/doi/10.1103/PhysRevB.102.155149}
}

@article{Goft2025,
      title={Engineering Topological Materials}, 
      author={Amit Goft and Eric Akkermans},
      year={2025},
      archivePrefix={arXiv},
    journal = {arXiv},
    pages = {2508.04927},
      primaryClass={cond-mat.mes-hall},
      url={https://arxiv.org/abs/2508.04927}, 
  doi = {https://arxiv.org/abs/2508.04927},
}

@article{Klitzing1980,
  title = {New Method for High-Accuracy Determination of the Fine-Structure Constant Based on Quantized Hall Resistance},
  author = {Klitzing, K. v. and Dorda, G. and Pepper, M.},
  journal = {Phys. Rev. Lett.},
  volume = {45},
  issue = {6},
  pages = {494--497},
  numpages = {0},
  year = {1980},
  month = {Aug},
  publisher = {American Physical Society},
  doi = {https://doi.org/10.1103/PhysRevLett.45.494},
  url = {https://link.aps.org/doi/10.1103/PhysRevLett.45.494}
}

@article{Hu2015,
  title = {Measurement of a Topological Edge Invariant in a Microwave Network},
  author = {Hu, Wenchao and Pillay, Jason C. and Wu, Kan and Pasek, Michael and Shum, Perry Ping and Chong, Y. D.},
  journal = {Phys. Rev. X},
  volume = {5},
  issue = {1},
  pages = {011012},
  numpages = {7},
  year = {2015},
  month = {Feb},
  publisher = {American Physical Society},
  doi = {https://doi.org/10.1103/PhysRevX.5.011012},
  url = {https://link.aps.org/doi/10.1103/PhysRevX.5.011012}
}

@article{mittal2016,
	title = {Measurement of topological invariants in a {2D} photonic system},
	volume = {10},
	issn = {1749-4893},
	url = {https://doi.org/10.1038/nphoton.2016.10},
	doi = {https://doi.org/10.1038/nphoton.2016.10},
	number = {3},
	journal = {Nature Photonics},
	author = {Mittal, Sunil and Ganeshan, Sriram and Fan, Jingyun and Vaezi, Abolhassan and Hafezi, Mohammad},
	month = mar,
	year = {2016},
	pages = {180--183},
}

@article{Thouless1982,
  title={Quantized Hall Conductance in a Two-Dimensional Periodic Potential},
  author={Thouless, D. J. and Kohmoto, M. and Nightingale, M. P. and den Nijs, M.},
  journal={Phys. Rev. Lett.},
  volume={49}, pages={405--408}, year={1982},
  doi={10.1103/PhysRevLett.49.405}
}

@article{Chalopin2020,
  author={Chalopin, T. and others},
  title={Probing chiral edge dynamics and bulk topology of a synthetic Hall system},
  journal={Nat. Phys.}, volume={16}, pages={1017--1021}, year={2020},
  doi={10.1038/s41567-020-0942-5}
}

@article{Jotzu2014,
  author={Jotzu, G. and others},
  title={Experimental realization of the topological {Haldane} model with ultracold fermions},
  journal={Nature}, volume={515}, pages={237--240}, year={2014},
  doi={10.1038/nature13915}
}

@article{Koenig2007,
  author={K{\"o}nig, M. and others},
  title={Quantum Spin {Hall} Insulator State in {HgTe} Quantum Wells},
  journal={Science}, volume={318}, pages={766--770}, year={2007},
  doi={10.1126/science.1148047}
}

@article{Niemi1984,
  author={Niemi, A. J. and Semenoff, G. W.},
  title={Spectral asymmetry on an open space},
  journal={Phys. Rev. D}, volume={30}, pages={809--818}, year={1984},
  doi={10.1103/PhysRevD.30.809}
}

@book{Katsnelson1979,
  author={Katsnelson, M. I.},
  title={Graphene: Carbon in Two Dimensions},
  publisher={Cambridge University Press}, address={Cambridge}, year={2012}
}

@article{Kane2005a,
  author={Kane, C. L. and Mele, E. J.},
  title={{$Z_2$} Topological Order and the Quantum Spin Hall Effect},
  journal={Phys. Rev. Lett.}, volume={95}, pages={146802}, year={2005},
  doi={10.1103/PhysRevLett.95.146802}
}

@article{Kane2005b,
  author={Kane, C. L. and Mele, E. J.},
  title={Quantum Spin Hall Effect in Graphene},
  journal={Phys. Rev. Lett.}, volume={95}, pages={226801}, year={2005},
  doi={10.1103/PhysRevLett.95.226801}
}

@article{Fu2006,
  author={Fu, L. and Kane, C. L.},
  title={Time reversal polarization and a {$Z_2$} adiabatic spin pump},
  journal={Phys. Rev. B}, volume={74}, pages={195312}, year={2006},
  doi={10.1103/PhysRevB.74.195312}
}

@article{Flynn1997,
  author={Flynn, D. P.},
  title={Two-dimensional phase unwrapping with minimum weighted discontinuity},
  journal={J. Opt. Soc. Am. A}, volume={14}, pages={2692--2701}, year={1997},
  doi={10.1364/JOSAA.14.002692}
}

@unpublished{Faktor2026,
  author={Faktor, S. and Abulafia, Y. and Akkermans, E.},
  title={Imaging Topology in Bipartite Lattices: Wavefunction Dislocations and Edge States},
  note={preprint (2026)}
}

@unpublished{Makwana2026,
  author={Makwana, V. and Akkermans, E.},
  title={Local Defects and the Topology of the {Haldane} Model},
  note={preprint (2026)}
}

@unpublished{Orion2026,
  author={Orion, N. and Rotstein, B. and Akkermans, E.},
  title={Topological Dislocations in Two-Qubit Systems},
  note={preprint (2026)}
}

@article{Abulafia2025,
  author={Y.~Abulafia and A.~Goft and N.~Orion and E.~Akkermans},
  title={Defects Potentials for Two-Dimensional Topological Materials},
  journal={Phys. Rev. B}, volume={113}, pages={045102}, year={2026},
  doi={10.1103/mw16-6bwh}
}

\end{document}